\documentclass[11pt,a4paper]{article}
\usepackage{jheppub}
\usepackage{amssymb}
\usepackage{graphicx}
\usepackage{latexsym}
\usepackage{amsmath}
\usepackage{amsfonts}

\def\bR{{\mathbb{R}}}

\title{Aspects of Three-dimensional Spin-4 Gravity}

\author{Hai Siong Tan}
\affiliation{Berkeley Center for Theoretical Physics and Department of Physics,\\
	University of California, Berkeley, CA 94720-7300}
\emailAdd{haisiong\_tan@berkeley.edu}

\abstract{
We discuss some interesting holographical aspects of three dimensional higher-spin gravity with a negative cosmological constant in the framework of $SL(4,\bR)\times SL(4,\bR)$ Chern-Simons theory. Using a recently found technique, we construct explicitly a solution that can be interpreted as spin-$4$ generalization of the BTZ solution, and demonstrate how $\mathcal{W}_4$ symmetry and the higher-spin Ward identities arise from the bulk equations of motion coupled to spin-3 and spin-4 currents. We match the eigenvalues of a Wilson loop along the time-like direction of the BTZ to that of the spin-4 solution, and show that this yields remarkably consistent gravitational thermodynamics for the latter. This furnishes an important, concrete supporting example for a recent proposal to understand spacetime geometries in three-dimensional higher-spin gravity formulated via $SL(N,\bR)\times SL(N,\bR)$ Chern-Simons theories.}


\begin{document}
\maketitle
\flushbottom

\renewcommand{\theequation}{\thesection.\arabic{equation}} \csname
@addtoreset\endcsname{equation}{section}
\def\d{\delta}
\def\D{\Delta}
\def\e{\epsilon}
\def\h{\eta}
\def\L{\Lambda}
\def\x{\xi}
\def\X{\Xi}
\def\p{\pi}
\def\P{\Pi}
\def\vp{\varpi}
\def\r{\rho}
\def\vr{\varrho}
\def\s{\sigma}
\def\S{\Sigma}
\def\t{\tau}
\def\u{\upsilon}
\def\U{\Upsilon}
\def\f{\phi}
\def\F{\Phi}
\def\vf{\varphi}
\def\c{\chi}
\def\ps{\psi}
\def\Ps{\Psi}
\def\o{\omega}
\def\O{\Omega}
\def\cA{{\cal A}}
\def\cB{{\cal B}}
\def\cC{{\cal C}}
\def\cD{{\cal D}}
\def\cE{{\cal E}}
\def\cF{{\cal F}}
\def\cG{{\cal G}}
\def\cH{{\cal H}}
\def\cI{{\cal I}}
\def\cJ{{\cal J}}
\def\cK{{\cal K}}
\def\cL{{\cal L}}
\def\cM{{\cal M}}
\def\cN{{\cal N}}
\def\cO{{\cal O}}
\def\cP{{\cal P}}
\def\cQ{{\cal Q}}
\def\cR{{\cal R}}
\def\cS{{\cal S}}
\def\cT{{\cal T}}
\def\cU{{\cal U}}
\def\cV{{\cal V}}
\def\cW{{\cal W}}
\def\cX{{\cal X}}
\def\cY{{\cal Y}}
\def\cZ{{\cal Z}}
\def\e{\epsilon}
\def\be{\begin{equation}}
\def\ee{\end{equation}}
\def\bea{\begin{eqnarray}}
\def\eea{\end{eqnarray}}
\def\ba{\begin{array}}
\def\ea{\end{array}}
\def\nn{\nonumber}
\def\tr{\text{tr}}
\def\dim{3}
\def\ww{\wedge}
\def\bra{\langle \,}
\def\sigmaet{\, \rangle}
\def\comma{\,,\,}
\newcommand\smargin[1]{\mbox{}\marginpar
{\raggedright\hspace{0pt}{\it\small S.T.:\,#1}}}

\section{Introduction}
\label{sec:Intro}

Theories of higher-spin gravity, as introduced in the seminal papers of Vasiliev and collaborators in \cite{vasiliev_int,Bekaert:2005vh}, have gathered a resurging interest recently due to their promising role within the context of the AdS/CFT correspondence. For example, in \cite{Douglas}, a certain higher spin gravity theory with $hs(d-1,2)$ algebra\footnote{This is a non-abelian higher-spin algebra that contains $o(d-1,2)$ as a subalgebra. We refer the reader to Section 5 of \cite{Bekaert:2005vh} for details.} appeared in the context of a constructive derivation of holography for free field theory in $d$-dimensions. In another well-known example, there is a conjectured duality \cite{Klebanov:2002ja,Sezgin} between Vasiliev's theory of higher spin gravity in $AdS_4$ and $O(N)$ vector models. Most recently, Gaberdiel and Gopakumar proposed in \cite{Gaberdiel:2010pz} that in a certain large-$N$ limit, $\cW_N$ minimal models are dual to a Vasiliev-type higher spin theory in $AdS_3$ coupled to two complex scalar fields. 

More precisely, in Gaberdiel-Gopakumar conjecture, the boundary CFT can be represented as a diagonal coset of WZW models of the form 
\be
\label{WNminimal}
\frac{SU(N)_k \oplus SU(N)_1}{SU(N)_{k+1}},
\ee
with the large-$N$ limit corresponding to taking
\be
\label{thooftlimit}
k,N\rightarrow \infty,\,\,\lambda = \frac{N}{k+N} \,\,\text{being fixed}
\ee
where $\lambda$ serves the role of a 't Hooft parameter.
This conjecture is very interesting since it is known that the $\cW_N$ CFT is integrable, and in principle, correlation functions can be computed precisely for all $N$ and $k$.\footnote{In \cite{Papadodimas:2011pf}, the three and four point functions in the minimal CFT were computed, and it was argued that there are several additional light states difficult to see in the bulk and which do not decouple (see also \cite{Chang:2011mz} for related issues). Most recently, in \cite{Castro}, it was argued that these light states can be identified with bulk solutions that represent higher-spin analogues of conical singularities after an appropriate analytic continuation.} It relies on the equivalence between the $\cW_N$ algebra and Vasiliev-type higher-spin algebra, and how the former dictates the representation theory of the minimal CFT in the limit above, as was recently explained in \cite{Gaberdiel:2011wb}. A non-trivial evidence for this conjecture also appeared recently in \cite{Gaberdiel:2011zw} where the one-loop determinant of the gravitational theory was shown to be precisely the vacuum character of $\cW_N$.

On another note, higher spin-$N$ theories in three dimensions are more manageable to work with than their higher-dimensional analogues since it is consistent to truncate the tower of higher spin fields to those with spin $s<N$ \cite{Blencowe}. The massless higher spin gauge fields possess no local degrees of freedom and can be regarded as higher spin versions of the graviton which is topological in three dimensions. The global degrees of freedom are those which are associated with boundary excitations of the fields, with the algebra of the asymptotic symmetry group being enlarged from two copies of the Virasoro algebra to two copies of the $\mathcal{W}_N$ algebra. They are formulated via $SL(N,\bR)\times SL(N,\bR)$ Chern-Simons theories. To make contact with Gaberdiel-Gopakumar conjecture discussed above, we need to take the large $N$ limit indicated in \eqref{WNminimal}, upon which we have $hs[\lambda]\oplus hs[\lambda]$ Chern-Simons theory, with $\cW_{\infty} [\lambda]$ as the asymptotic algebra \cite{Henneaux:2010xg}. 

In \cite{Kraus1,Kraus2,Maloney}, the authors presented geometries that were argued to be generalized BTZ solutions that carry spin-three charges. The spacetime metric of these solutions describe a traversable wormhole connecting two asymptotic regions, but under a higher spin transformation found in \cite{Kraus2}, these solutions then describe black holes with manifestly smooth event horizons. The authors then argued that a gauge-invariant characterization of a smooth horizon for any solution of the $SL(3,\bR) \times SL(3,\bR)$ theory lies in \emph{matching the eigenvalues} of the Wilson holonomy along the time-like direction to that of the BTZ. This was shown to yield consistent gravitational thermodynamics for the solutions.

Apart from being interesting in its own right, this class of results turns out to have some interesting implications for the Gaberdiel-Gopakumar conjecture. When these solutions are lifted to $hs[\lambda]\oplus hs[\lambda]$ by adding an infinite series of higher-spin charges and appropriately replacing ordinary multiplication with the lone-star product, it was demonstrated remarkably in \cite{Kraus3} that this $hs[\lambda]$ solution yields a high-temperature partition function that agrees with that of the boundary CFT at $\lambda = 0,1$, with spin-3 chemical potential inserted. It was argued that the partition function in this limit is shared by the coset minimal model in Gaberdiel-Gopakumar conjecture since these Chern-Simons solutions describe the topological sector of the bulk, and that the results should support the conjecture for other values of $\lambda$ too.

Against this backdrop, the main purpose of this paper is to study the elegant program of \cite{Kraus1} more  concretely in the case of $N=4$. By our current level of understanding, the consistency of such a holonomy prescription cannot be guaranteed by asymptotic symmetry arguments alone, and it would be important to investigate some manageable cases. We will see that the spin-4 solution furnishes a non-trivial supporting example of various aspects of this proposal. Also, our solution can be used as a useful background limit, when the higher spin currents are turned on in the $hs[\lambda]$ case\cite{Kraus3}. 

The outline of this paper is as follows: in Section 2, we discuss the basic formulation of higher-spin gravity in the framework of Chern-Simons theory, in particular, for the case of spin-4. (This was done for the case of spin-3 in \cite{Camp:2010}.) Then, we  demonstrate how $\cW_4$ symmetry and the higher-spin Ward identities arise from the bulk equations of motion coupled to spin-3 and spin-4 currents\footnote{Recently, in an elegant paper \cite{Camp:2011}, the authors provide a closed formula for the structure constants of all classical $\mathcal{W}_N$ algebras. Their approach relies on obtaining the algebras from the Poisson brackets of the charges that generate these transformations. For us, following \cite{Kraus1, Kraus2}, we translate these variations into OPEs for the symmetry currents. The two approaches are equivalent.}, and briefly discuss the non-principal embeddings of $SL(2,\bR)$ in $SL(4,\bR)$ with their associated Chern-Simons vacua. 
In Section 3, we write down explicitly the solution that can be interpreted as the spin-4 generalization of the BTZ via the techniques introduced in \cite{Kraus1}, and demonstrate how the holonomy prescription yields a remarkably consistent gravitational thermodynamics in this case. Finally, we end off with a summary of our results and a few suggestions for future work. Appendix A collects our conventions for the $SL(4,\bR)$ generators, while Appendix B contains the explicit field equations of motion. 

\section{Spin-4 gravity, $SL(4,\bR)\times SL(4,\bR)$ Chern-Simons, and $AdS_3$ vacua}
\label{sec:Prelim}
\subsection{The basic formulation}
Let us begin by reviewing the basic formulation of higher-spin gravity in the framework of Chern-Simons theory. Recall that the Chern-Simons action reads
\be 
\label{CSaction}
S_{CS} [A]=\frac{k}{4\pi} \int \text{tr} \left( A\wedge dA + \frac{2}{3} A\wedge A \wedge A \right)\,.
\ee
In a remarkable observation by Witten in \cite{Witten}\footnote{See also \cite{Townsend}.}, it was noted that the combination (with the same Chern-Simons level $k$)
\be 
\label{action}
S = S_{CS}[A] - S_{CS}[\widetilde{A}]
\ee
where $A$ and $\widetilde{A}$ are independent Chern-Simons connections labelled in $SL(2,\bR)$, reduces to the Einstein-Hilbert action\footnote{We note in passing that if the Chern-Simons levels are allowed to be different, we then have topologically massive gravity\cite{Deser:1981wh}, of which a higher-spin analogue was considered most recently in \cite{Chen:2011yx}.} if we identify
\be 
\label{vector1}
A=\left( \omega^a + \frac{e^a}{l} \right)J_a\,,\qquad \tilde{A}=\left( \omega^a - \frac{e^a}{l} \right)J_a
\ee
where the one-forms $e^a, \omega^a$ are the vielbeins and spin connection, and $J_a$ are the $SL(2,\bR)$ generators. This identification is up to boundary terms, and, in particular, is made with the normalization $\text{Tr}(J_aJ_b) = \frac{1}{2}\, \h_{ab}$, and the identification $k=\frac{l}{4 G}$. This can be generalized to an $SL(N,\bR)\times SL(N,\bR)$ Chern-Simons action, with the vector potential expressed as  
\be 
\label{vector}
A=\left( \omega^a + \frac{e^a}{l} \right)J_a +  \sum_{i=2}^{N-1} \left( \omega^{a_1a_2\ldots a_i} + \frac{e^{a_1a_2\ldots a_i}}{l} \right)T_{a_1a_2\ldots a_i}
\ee
where $e^{a_1a_2\ldots a_i}, \omega^{a_1a_2\ldots a_i}$  are the analogous gauge potentials for the higher-spin fields, and $T_{a_1a_2\ldots a_m}$ are the spin-$m$ generators which are completely symmetric and traceless in their indices (i.e. $T^{b}_{ba_3\ldots a_i}=0$). Like \eqref{vector1}, the expression for $\widetilde{A}$ is similar but with $e\rightarrow -e$.

The higher-spin generators satisfy 
\be 
\label{generator expression}
\left[J_a\,,\,J_b\right]=\epsilon_{abc}J^c,\qquad \left[J_b\,,\,T_{a_1a_2\ldots a_{s-1}} \right] = {\epsilon^m}_{b(a_1}T_{a_2\ldots a_{s-1})m}\,,
\ee
and they clearly transform as $SL(2,\bR)$ tensors. For a general $N$, when the vector potentials are valued as in \eqref{vector}, we have a consistent description of a `gravitational' sector. More precisely, when the equations of motion are linearized, we obtain the physics of a spin-$N$ field propagating on an $AdS_3$ background (see, for example, Section 2 of \cite{Camp:2010} for a brief review). In Appendix B, we write down the field equations of motion explicitly. From them, it is straightforward to check that apart from the usual diffeomorphism, the spin-2 (and the higher-spin) fields acquire new gauge transformations proportional to the spin-3 gauge parameters and spin-4 gauge parameters. 

An useful basis which we will rely on in later sections is one in which does not take into
account the trace constraints on the generators. One general expression for such a basis for higher-spin fields (see, for example, \cite{Camp:2011}) is
\bea
\label{basic}
[L_+\,,L_-]&=&2L_0,\,\,\, [L_{\pm}\,,L_0]=\pm L_{\pm},\nn \\
\left[L_i\, ,W^l_m\right] &=&\left(il-m\right) W^l_{i+m}, \nn \\
W^l_m&=&(-1)^{l-m} \frac{(l+m)!}{(2l)!}   ad_{L_-}^{l-m} \left( L^l_+ \right)\,.
\eea
In the above notations, the spin is $(l+1)$, $i=0,\pm 1$, $-l\leq m\leq l$, and $ad_L(f)=[L,f]$ refers to the adjoint action of $L$ on $f$. This is easily motivated by letting $W^l_l=L_+^l$ in the fundamental, and then deriving the rest of the generators by the lowering operator $L_-$. In \cite{Camp:2010}, the isomorphism in the case of spin-3 between the spin generators $T_{a_1a_2\ldots a_{s-1}}$ and the $W^l_m$ generators is computed. Similarly, we derive the isomorphism in the spin-4 case. Up to one constant scaling factor, we find that for the ten spin-4 generators, the mapping goes as
\begin{align}
&T_{222}= U_0,\,\,T_{220}=\frac{1}{2} \left( U_1 + U_{-1} \right),\,\,T_{221} = \frac{1}{2} \left( U_1 - U_{-1} \right),\,\,T_{200}=\frac{1}{4} \left( U_2 + U_{-2} \right) + \frac{1}{2} U_0, \nn \\
&T_{012} = \frac{1}{4} \left( U_2 - U_{-2} \right),\,\,T_{211}=\frac{1}{4} \left( U_2 + U_{-2} \right)-\frac{1}{2}U_0, T_{000}=\frac{1}{8} \left( U_3 + U_{-3} + 3(U_1+U_{-1}) \right),\nn \\
&T_{001}=\frac{1}{8} \left( U_3 - U_{-3} + U_1 - U_{-1} \right),\,\,T_{011}=\frac{1}{8} \left( U_3 + U_{-3} - U_1 - U_{-1} \right),\nn \\
&T_{111}=\frac{1}{8} \left( U_3 - U_{-3} + 3(U_{-1} - U_1) \right)
\end{align}
where we have denoted $\cU_m = \cW^3_m$. For the general spin-$N$ case, we can derive this isomorphism straightforwardly by starting with the `highest-weight'  generator $T_{22\ldots 2}$. 

The physical interpretation of these Chern-Simons theories begins after we identify metric-like fields. 
They are identified by demanding invariance under local Lorentz invariance. This condition is only sufficient for the spin-2 (metric) and spin-3 fields. In \cite{Camp:2011}, it was explained that for higher-spin fields, the choice of identification is unique for spin-4 and spin-5 fields if we further demand that in the linearized regime, rewriting them in terms of vielbeins reproduces the definition in the free theory. Then, it can be shown (see \cite{Camp:2010, Camp:2011}) that this leads to the following definitions of metric-like fields:
\be
\label{field definitions}
g \sim \textrm{tr} (e\cdot e),\qquad \psi_3 \sim \textrm{tr} (e\cdot e \cdot e),\qquad  \psi_4 \sim \textrm{tr}(e^4) - \frac{3\lambda^2-7}{10} \left( \textrm{tr} e^2\right)^2
\ee
where `tr' in \eqref{field definitions} is defined\footnote{This is essentially a bilinear invariant form on $hs[\lambda]$. We refer the reader to Section 3.1 of \cite{Camp:2011} for details. Please note that `Tr' is used to denote taking the matrix trace in our paper.}  
via equation (3.14) of \cite{Camp:2011}, and
$\lambda$ is related to the quadratic Casimir by
\be
\label{casimir}
L_0^2 - \frac{1}{2} \left( L_+ L_- + L_-L_+ \right) = \frac{\lambda^2-1}{4} .
\ee
In our case, $\lambda = 4$, but is not necessarily an integer if we consider the larger framework of $hs[\lambda]$ algebra. Consider building an infinite tower of higher-spin $(l+1)$ fields in the basis of generators $W_m^l$, each appearing once, so $l$ in \eqref{basic} runs from 1 to $\infty$. In general this algebra $hs[\lambda]= \oplus_{l=1}^{\infty} \mathfrak{g}^{(l)}$ (where $\lambda$ is as defined in \eqref{casimir}) is distinct for different values of $\lambda$, and when $\lambda=N \in \mathbb{Z}$, all higher-spin generators for spin $>N$ can be truncated and the algebra reduces to $SL(N,\bR)$ algebra. An interesting fact is that the commutator between even-$l$ $W$'s yields a sum of odd-$l$ $W$'s, whereas the commutator between even-$l$ $W$'s and odd-$l$ $W$'s yields a sum of even-$l$ $W$'s. A related implication of this is that it is possible for the algebra to be truncated of all even-$l$ generators (which correspond to odd spins). In four dimensions, the well-known counterpart is the minimal Vasiliev model. 

\subsection{$\cW_4$ symmetry and OPEs from field equations}

We first summarize how $\mathcal{W}_N$-algebras emerge from the asymptotic symmetries of $AdS_3$, and certain holographic aspects of the boundary CFT with $\mathcal{W}_N$ symmetry. Let us begin with the Fefferman-Graham expansion in pure three-dimensional gravity that parametrizes asymptotically $AdS_3$ solutions with a flat boundary metric (see \cite{review_banados}):
\be \label{sol_einstein}
ds^2 = \, l^2 \left\{\, d\rho^2 - \frac{8\pi G}{l} \left(\, \cL\, (dx^+)^2 + \widetilde{\cL}\, (dx^-)^2  \,\right) - \left(\, e^{2\rho} + \frac{64\pi^2G^2}{l^2}\,\cL\,\widetilde{\cL}\, e^{-2\rho}  \,\right) dx^+ dx^- \,\right\}
\ee
where $(\rho,x^{\pm}\equiv t\pm \phi)$ describes the solid cylinder, and $\cL=\cL(x^+),\,\tilde{\cL} = \tilde{\cL}(x^-)$ are arbitrary functions of $x^{\pm}$\footnote{For example, $\cL = \tilde{\cL} = -\frac{M}{4\pi}$ for the static BTZ of mass $M$, with the global $AdS_3$ vacuum corresponding to $M = -1/8G$. If $\cL=\tilde{\cL}=0$, we recover the Poincar\'e patch of $AdS_3$.}. In terms of the Chern-Simons connections, denoting $b=e^{\rho L_0}$,
\bea
\label{ads3}
A&=& b^{-1} a\left(x^{+} \right) b +b^{-1} db,\,\,\, \bar{A}=b\bar{a}\left(x^-\right)b^{-1} + bdb^{-1}, \nn \\
a&=&\left(  L_1 - \frac{2\pi}{k} \cL L_{-1}  \right) dx^+,\,\,\,\bar{a}=\left( -L_{-1} + \frac{2\pi}{k} \tilde{\cL} L_1 \right)dx^-
\eea
It was argued in \cite{Camp:2010} that in the context of this higher-spin theory, a more appropriate definition of an asymptotically $AdS_3$ solution is the set of conditions:
\be 
\label{bc}
\left( A - A_{AdS_3} \right)  \Big |_{\textrm{boundary}} = \mathcal{O}(1),\,\,\,A_{\rho}=L_0,\,\,\, A_- = 0
\ee
where the first equation in \eqref{bc} refers to a finite difference at the boundary (as $\rho\rightarrow \infty$). Similar expressions hold for the anti-holomorphic sector. It was then explained in \cite{Camp:2010,Camp:2011} that \eqref{bc} translates into the Drinfeld-Sokolov condition on $A$, and if we consider the branching of $\mathfrak{g}$ according to the sign of the eigenvalues of the adjoint action of $L_0$, this implies, by \eqref{basic} that we can set terms in $W^l_m$ where $m$ is positive to vanish. These are first-class constraints which generate gauge transformations, and we can fix the residual gauge freedom by letting only terms in $W^l_{-l}$ to survive.\footnote{This procedure is the Drinfeld-Sokolov reduction in the highest-weight gauge. See, for example,\cite{deBoer1} and \cite{Bouwknegt:1992wg}.} A similar procedure works for the anti-holomorphic $\bar{A}$. Altogether, we end up with the ansatz for the connections:
\be
\label{gads3}
a = \left( L_1 + \sum_{l} \cW^l_{-l} W^l_{-l}    \right) dx^+,\qquad \bar{a} = -\left( L_{-1} + \sum_{l} \bar{\cW}^l_{l} W^l_{l}    \right) dx^-
\ee
with the $\cW,\bar{\cW}$'s being general functions of $\phi$, and $A,\bar{A}$ obtained by gauge transforming via $b$ as in \eqref{ads3}. The global symmetries of the space of solutions described by $a$ are described by the gauge transformations 
\be
\label{globalgauge}
\lambda(\phi)=\sum_i \xi^i(\phi) L_i + \sum_{l,m} \chi^l_m (\phi) W^l_m\,.
\ee
Identifying those that leave the structure of \eqref{gads3} invariant, we can express each gauge parameter $\chi^l_m, m<l$ as functions of the fields $\cW^l_{-l}$, $\chi^l_l$ and their derivatives. Finally, we can write down the gauge transformations $\delta_{\chi} \cW$ of $\cW^l_{-l}$ with respect to the parameters $\chi^l_l$. In \cite{Camp:2010, Camp:2011}, from this point, the asymptotic symmetry algebra is then obtained from the Poisson brackets of the charges that generate these transformations, and we obtain the two copies of $\cW_N$-algebra. 

Now in \cite{Kraus1, Kraus2}, a slightly different approach was adopted to elucidate both the emergence of the $\cW_N$-algebras and some holographic aspects at the same time. It was shown, explicitly in the case of spin-3, that the bulk field equations evaluated on a more general ansatz than \eqref{gads3} (which corresponds to generalized boundary conditions) yields the Ward identities in the CFT in the presence of spin-3 sources. This means that certain terms in the connection can be related precisely to extra source terms in the boundary CFT lagrangian, making feasible the existence of an AdS/CFT dictionary for the higher-spin sources. This was argued to be important in demonstrating that we have a consistent holographical dictionary for computing correlation functions of the stress tensor and spin-3 currents. By invoking Noether's theorem
\be
\label{Noether}
\delta \cO = 2\pi \textrm{Res}_{z\rightarrow 0} \left[ \sum_l \chi^l_l(z) \cW^l_{-l}(z) \cO(0)   \right]\,,
\ee
upon obtaining the various $\delta \cW$, one can also read off the OPEs between the Virasoro primary fields of the $\cW_N$ algebra conveniently. In \cite{Kraus2}, this was done for the $\cW^{(2)}_3$ algebra.

One of the main purposes of this work is to study explicitly if the procedures developed for the spin-3 case in \cite{Kraus1, Kraus2} generalize neatly to the spin-4 case as well. In this Section, we shall perform an analogous calculation for the spin-4 case below. 

We begin with the ansatz
\bea
\label{spin4ansatz}
&a=\left( L_1+ \alpha \cL L_{-1} + \beta \cW W_{-2} + \gamma \mathcal{U} U_{-3}  \right) dx^+
+\left( \sum_{m=-2}^2 \chi_m W_m  + \sum_{m=-3}^3 f_m U_m + \nu L_{-1}  \right) dx^- \nn \\
&\bar{a}=-\left( L_{-1}+ \alpha \bar{\cL} L_{1} + \beta \bar{\cW} W_{2} + \gamma \bar{\mathcal{U}} U_{3}  \right) dx^-
-\left( \sum_{m=-2}^2 \bar{\chi_m} W_{-m}  + \sum_{m=-3}^3 \bar{f}_m U_{-m} + \bar{\nu} L_{1}  \right) dx^+\,, \nn\\
\eea
where $W$'s are the spin-3 generators, $U$'s are the spin-4 ones, all fields are functions of $x^{\pm}$, and $\{ \alpha, \beta, \gamma \}$ are scaling parameters to be specified later. 

In the language of Chern-Simons theory, the bulk field equations are conditions for flat connections. From the viewpoint of the boundary CFT with $\cW_4$ symmetry, the fields $\cW$ and $\mathcal{U}$ are dimension $(3,0)$ and $(4,0)$ primary fields, the anti-holomorphic ones being $(0,3)$ and $(0,4)$ primaries. The ansatz \eqref{spin4ansatz} generalizes the one in \eqref{gads3} and the modified boundary conditions can be interpreted as adding to the boundary CFT sources terms 
\be
I \rightarrow I - \int d^2 x \left( \chi_2(x) \mathcal{W}(x) +  \bar{\chi_2}(x) \bar{\mathcal{W}}(x) + f_3(x) \mathcal{U}(x) + \bar{f_3}(x) \bar{\mathcal{U}}(x) \right)
\ee
It turns out that the functions $\chi_2, \bar{\chi}_2, f_3, \bar{f}_3$ can be identitifed with those in the connections \eqref{spin4ansatz} by imposing the bulk field equations to yield the Ward identities of the CFT in the presence of these spin-3 and spin-4 sources.  As explained in \cite{Kraus1}, the higher-spin operators are irrelevant in the RG sense and adding them changes the UV structure of the CFT. Correspondingly, the bulk geometry will asymptote to a different $AdS_3$ geometry. We will discuss this more explicitly later on in Section 2.3. 

Below, we analyze the holomorphic connection $a$ which is gauge-equivalent to $A$.\footnote{The corresponding computation for $\bar{A}$ is similar.} The flat connection condition $da + a\wedge a = 0$ gives rise to the following system of differential equations for sixteen unspecified functions of $x^{\pm}$. There are fourteen equations with two free parameters $\chi_2, f_3$ that can be chosen freely. 
For notational simplicity, we denote $\chi_2 \equiv \chi, f_3 \equiv f$, and superscripted primes denote $\partial_{x^+}$. 
For the $\chi$'s:
\begin{align}
\chi_1&= - \chi' \nn \\
\chi_0&=2\alpha \cL \chi - 6 \beta f\cW + \frac{1}{2} \chi'' \nn \\
\chi_{-1} &= 4 \beta \cW f' - \frac{2}{3} \alpha \chi \cL' + 2\beta f \cW'  - \frac{5}{3} \alpha \cL\chi' - \frac{1}{6}\chi''' \nn \\
\chi_{-2} &= \alpha^2 \cL^2 \chi -3 \gamma \mathcal{U} \chi -\frac{3}{2} \beta f' \cW'+\frac{7}{12} \alpha \cL' \chi' \nn \\
&\,\,-\frac{13}{10} \beta \cW f''+\frac{1}{6} \alpha \chi \cL'' -\frac{1}{10} f \left(48 \alpha \beta \cL \cW+5 \beta \cW'' \right)+\frac{2}{3} \alpha \cL \chi''+\frac{1}{24} \chi^{(4)} \nn \\
\end{align}
while for the $f$'s, we have
\begin{align}
&f_2= - f' \nn \\
&f_1 = \frac{1}{2} \left( 6\alpha f \cL + f''  \right) \nn \\
&f_0 = -\frac{8}{3} \alpha \cL f' - \alpha f  \cL' - \frac{1}{6} f''' \nn \\
&f_{-1}= \frac{1}{24} \left(48 \beta \cW \chi+22\alpha f' \cL'+28 \alpha \cL  f''+6 f \left(12  \alpha^2 \cL^2+12 \gamma \cU+ \alpha \cL'' \right)+f^{(4)}\right) \nn \\
&f_{-2} = \frac{1}{120} \Bigg(-264 \alpha^2 \cL^2 f'-216 \gamma \cU f'-72 \gamma f \cU'-48 \beta \chi  \cW' -192 \beta \cW \chi'  \nn \\ 
&\qquad -50 \alpha \cL' f''  - 28 \alpha f' \cL'' - 8 \alpha \cL \left(27 \alpha f \cL' + 5f''' \right)-6\alpha f \cL''' - f^{(5)} \Bigg) \nn \\
&f_{-3} = \frac{1}{720} \Bigg( 288 \gamma f' \mathcal{U}'+240 \beta \cW' \chi'+544 \alpha^2 \cL^2 f''+360 \gamma \mathcal{U} f''+78 \alpha f'' \cL''+48 \beta \chi \cW'' \nn \\
&\qquad+432 \beta \cW \chi''+90 \alpha \cL' f''' +34 \alpha f' \cL''' +2 \alpha\cL \left(720 \beta \cW \chi+482 \alpha f' \cL'+25 f''''\right)+6 f (120 \alpha^3 \cL^3 \nn \\
&\qquad-480 \beta^2 \cW^2+36 \alpha^2 \cL'^2+ \alpha \cL \left(264 \gamma \mathcal{U}+46 \alpha \cL''\right) +12 \gamma \mathcal{U}''+ \alpha\cL^{(4)})+f^{(6)}\Bigg) 
\end{align}

Finally, the higher spin-fields and $\nu$ are subject to the following equations:
\bea
\label{eqnfornu}
\nu &=& \frac{6}{5} \left( 9\gamma \mathcal{U} f - 4 \beta \cW \chi \right) \\
\label{eqnforL}
\alpha \partial_- \cL &=& \nu' + \frac{12}{5} \beta \cW \chi' - \frac{18}{5} \gamma \mathcal{U} f' \\
\label{eqnforW}
\beta \partial_- \cW &=& 
-3 \gamma \left[
     \mathcal{U}' \chi + 2\mathcal{U}\,\chi' \,\right] \,+\, \frac{\alpha}{12} \left[2\cL'''\chi + 9\cL''\chi' 
       +15\cL' \chi''  +10\cL\,\chi''' \,\right] \nn \\
&& +\, \frac{8\alpha^2}{3} \left[\cL \cL'\chi+ \cL^2 \chi' \,\right]
     \,+\, \frac{1}{24}\, \chi^{(5)} -\left( \frac{24}{5}\alpha \beta \cL' \cW + \frac{34}{5}\alpha \beta\cL\cW' + \frac{1}{2}\beta\cW''' \right)f \nn \\
&& - \left( \frac{44}{5}\alpha \beta \cL \cW + 2 \beta\cW''   \right)f' - \frac{14}{5}\beta\cW' f'' - \frac{13}{10} 
\beta\cW f''' \nn \\ 
\eea
\bea
\label{eqnforU}
\gamma \partial_- \mathcal{U} &=&  
\, \frac{\beta}{15} \left[\cW'''\,\chi + 6\,\cW''\,\chi'+ 14\cW'\,\chi''
    +14\cW\, \chi''' \,\right] \nn \\
&&+\frac{2\alpha\beta}{15}\left[25\cL' \cW\chi
    + 18\cL \cW'\,\chi + 52\cL \cW \,\chi'  \,\right] \, \nn \\
&&+\, \frac{\gamma}{10} \left[\mathcal{U}'''\, f  
    + 5\mathcal{U}''\, f'
    + 9\mathcal{U}\, f''  + 6\mathcal{U}\, f'''  \,\right] \nn \\
&&+\frac{\alpha}{360}\left[\, 3 \cL^{(5)} f  
    + 20\cL^{(4)} f' +56\,  \cL''' f''
    +  84 \cL'' f''' + 70 \cL' f^{(4)} 
    +28\cL\, f^{(5)} \right]\nn \\
&& -12\beta^2 \left[\cW \cW'  f + \cW^2  f' \,\right] 
+ \frac{14}{5} \left[ \alpha\cL'\, \gamma\mathcal{U}f + \alpha\cL\, \gamma\mathcal{U}'\,f + 2\, \alpha\cL\, \gamma\mathcal{U}\,f' \,\right]  \nn\\
&& +\frac{\alpha^2}{180}
  \left[177 \,  \cL' \cL''f  + 78\cL \cL''' f 
    + 295\cL'^2 f'+\, 352\cL  \cL''f'+ 588\cL  \cL' f''     + 196 \cL^2 f'''\right] \nn\\  
&& + \frac{8\alpha^3}{5} 
  \left[\, 3\, \cL^2 \cL' f   + 2\,  \cL^3 f' \right]
  + \frac{1}{720} f^{(7)}  . 
\eea
We now come to an important point(the logic here parallels the spin-3 analysis in \cite{Kraus1}): adding the source terms to the CFT action causes the stress tensor to attain $\bar{z}$ dependence. Upon inserting 
\be
\label{boundaryinsert}
e^{ \int d^2 x \left( \chi_2(x) \mathcal{W}(x) +  \bar{\chi_2}(x) \bar{\mathcal{W}}(x) + f_3(x) \mathcal{U}(x) + \bar{f_3}(x) \bar{\mathcal{U}}(x) \right)}
\ee
within the expectation value of the stress-energy tensor, and invoking the OPEs between the stress-energy tensor and higher spin operators which read
\be
T(z)\cW(0) \sim \frac{3}{z^2} \cW(0) + \frac{1}{z} \partial\cW(0) + \mathcal{O}(1),\qquad 
T(z)\mathcal{U}(0) \sim \frac{4}{z^2} \mathcal{U}(0) + \frac{1}{z} \partial\mathcal{U}(0) + \mathcal{O}(1),\qquad   
\ee
we obtain
\be
\label{partialT}
\frac{1}{2\pi}\partial_{\bar{z}} \langle T(z,\bar{z}) \rangle_{\chi,f} = 2\cW' \chi + 3\cW \chi' + 3\mathcal{U}' f + 4\mathcal{U} f'
\ee
where $\langle \ldots \rangle$ denotes inserting \eqref{boundaryinsert} within the expectation value. Note that in obtaining \eqref{partialT}, we have expanded in powers of $\chi,\,f$ and invoke the useful formula $\partial_{\bar{z}}(1/z) = 2\pi \delta^{(2)}(z,\bar{z})$. Now, observe that if we set
\be
\label{scalingconstants1}
\beta=-\frac{5}{12} \alpha,\qquad \gamma =\frac{5}{18}\alpha
\ee
then upon setting
\be
\label{energytensor}
2\pi \cL = T
\ee
we find that \eqref{partialT} is nothing but \eqref{eqnforL}. Thus, the stress-energy tensor corresponds to one single term in the connection. This was observed for the spin-3 case in \cite{Kraus1}, and it is nice to see explicitly that it is true for the spin-4 case as well.  To fix $\alpha$, we note that in the absence of all the higher-spin charges and conjugate potentials, if we demand the solution to reduce to BTZ in the chart \eqref{sol_einstein}, then 
\be
\label{alpha}
\alpha=\frac{2\pi}{k},
\ee
which can be checked to yield precisely the Brown-Henneaux central charge $c=6k$. The normalization of the parameters in \eqref{scalingconstants1} will enter into the geometries of the solutions in Section 3. 

Apart from the stress-energy tensor Ward identity, we can also use the bulk equations and \eqref{Noether} to read off the higher-spin OPEs or Ward identities. As a concrete example, we can easily read off the OPEs between two spin-3 and two spin-4 currents in our normalization. From \eqref{eqnforW}, \eqref{eqnforU} and \eqref{Noether} (letting $\mathcal{O}=\cW$ and $\cU$), we have (suppressing the scaling constants to compare with existing results in literature, see for example \cite{Camp:2011} )
\be
\label{OPEforW}
2\pi\langle \cW(z) \cW(0) \rangle = \frac{5}{z^6} + \frac{5\cL}{z^4} + \frac{5 \cL'}{2z^3} +\frac{\left(\frac{8}{3}\cL^2 + \frac{3}{4} \cL'' -6\cU \right)}{z^2}+ \frac{ \left( -3\cU' + \frac{1}{6} \cL''' + \frac{8}{3} \cL \cL'     \right)}{z}
\ee
\bea
\label{OPEforU}
2\pi\langle \cU(z) \cU(0) \rangle &=&\frac{1}{z} \left( \frac{1}{120}\cL^{(5)} - 12\cW \cW' + \frac{14}{5} (\cU \cL)'  + \frac{1}{10} \cU''' + \frac{177}{180} \cL' \cL'' + \frac{13}{30} \cL \cL''' + \frac{24}{5} \cL^2 \cL'                                       \right) \nn \\
&&+ \frac{1}{z^2} \left( \frac{1}{18} \cL^{(4)} + \frac{1}{2} \cU'' - 12\cW^2 + \frac{28}{5}\cL \cU + \frac{59}{36}\cL'^2 + \frac{88}{45} \cL \cL'' + \frac{16}{5} \cL^3   \right) \nn \\
&&+ \frac{1}{5z^3} \left( 9\cU + \frac{14}{9} \cL''' + \frac{98}{3} \cL \cL'   \right) + \frac{6}{5z^4} \left( 3\cU + \frac{7}{6} \cL'' + \frac{49}{9} \cL^2 \right) \nn \\
&&+ \frac{14\cL' }{3z^5} + \frac{28\cL}{3z^6} + \frac{7}{z^8}
\eea
The above results provide a foothold for understanding the holographic dictionary in the presence of spin-3 and spin-4 sources, and is essentially, the spin-4 generalization of what was achieved in \cite{Kraus1}.

\subsection{Non-principal embeddings and their Chern-Simons vacua}
\label{sec:otherW4}

Before we proceed to discuss the spacetime interpretation of the ansatz \eqref{spin4ansatz}, let us mention that there are other Drinfeld-Sokolov procedures with which one can construct $\cW$-algebras. What we have done above corresponds to the choice of the ``principal embedding" of $SL(2,\bR)$ in $SL(4,\bR)$. Let us first quickly review the meaning of having different embeddings of $SL(2,\bR)$ in the general $SL(N,\bR)$ theory. 

As explained in for example \cite{deBoer1}, that one considers $SL(2,\bR)$ embeddings is closely related to the requirement that one wants the algebra to be an extended conformal algebra, i.e. containing the Virasoro as a subalgebra and other generators to be primary fields with respect to this Virasoro algebra.  To each $SL(2,\bR)$ embedding within the simple Lie algebra that underlies the affine algebra, one can associate a generalized classical Drinfeld-Sokolov reduction of the affine algebra to obtain a $\cW$-algebra. 

For $SL(N,\bR)$, the number of inequivalent $SL(2,\bR)$ embeddings is equal to the number of partitions of $N$, and the standard reduction leading to $\cW_N$ algebras is associated with the so-called principal embedding. Also, the inequivalent $SL(2,\bR)$ embeddings are completely characterized by the branching rules of the fundamental representation. In our discussion below where we will give explicit examples of the statements above, we will parametrize the branching by various $SL(2,\bR)$-multiplets, and ``spin" in this context refers to the dimensionality of the representation. The conformal weight of each field is obtained from the $SL(2,\bR)$ spin by adding one. 

What is the relevance of these non-principal embeddings in a three-dimensional $SL(N,\bR)\times SL(N,\bR)$ higher-spin theory? A nice discussion was first made in \cite{Kraus2}, where it was pointed out that non-principal embeddings describe $AdS_3$ vacua of possibly different radii, with the corresponding $\cW$-algebra as the asymptotic symmetry algebra. Specifically in the case of $N=3$, the Polyakov-Bershadsky algebra $\cW_3^{(2)}$ is the only non-principal embedding in $SL(3,\bR)$. A solution that represents an interpolation between the $\cW_3^{(2)}$ (in the UV) and $\cW_3$ vacua (in the IR) was constructed, and an elegant linearized analysis of the RG flow background was presented in \cite{Kraus2}.

Let us very briefly review the $N=3$ case as presented in \cite{Kraus2}. For the principal embedding $(\cW_3)$, we have one spin-1 multiplet generated by $(L_0, L_{\pm 1})$ and one spin-2 multiplet generated by $(W_0, W_{\pm 1}, W_{\pm 2})$. For $\cW_3^{(2)}$, the branching reads as: (i)one spin-1 multiplet $\left( \frac{1}{4}W_2, \frac{1}{2}L_0, -\frac{1}{4}W_{-2} \right)$, (ii)one spin-0 multiplet $\left( W_0 \right)$, (iii)two spin-1/2 multiplets: $\left( W_1, L_{-1} \right), \left( L_1, W_{-1} \right)$, where the  $\cW_3^{(2)}$'s generators have been expressed in terms of the $\cW_3$'s. An analysis similar to that done for $\cW_3$ can be done to obtain the classical  $\cW_3^{(2)}$ algebra. 

In terms of the Chern-Simons connections, we have
\bea
\label{w32}
A_{AdS_3} &=& e^{\rho} \left( \frac{1}{4}W_2  \right) dx^+ + \left(  \frac{1}{2}L_0   \right) d\rho, \nn \\
\bar{A}_{AdS_3} &=& -e^{\rho} \left( \frac{1}{4}W_{-2}  \right) dx^- - \left( \frac{1}{2}L_0  \right) d\rho
\eea
which translates to the metric (the higher spin field $\psi_{abc}=0$)
\be
\label{w32metric}
ds^2 = \frac{l^2}{4} \left( d\rho^2 - e^{2\rho} dx^+ dx^- \right)
\ee
which is $AdS_3$ with radius $=\frac{l}{2}$, if we assume the same metric normalization as that for the principal embedding. 

The difference in the $AdS$ radius can be traced to the trace relations of the new $SL(2,\bR)$ generators, in particular that of $L_0$ which is also known as the defining vector of the embedding. For $\cW^{(2)}_3$, we note that $\textrm{Tr} \left( \left( \frac{1}{2}L_0 \right)^2 \right) =  \frac{1}{4} \textrm{Tr} \left(  L_0^2 \right)$, giving rise to an $AdS_3$ of half the radius of that of the principal embedding. This implies that the overall normalization of the Chern-Simons action restricted to the $SL(2,\bR)$ subalgebra must be reduced by an overall factor of $1/4$. For a fixed Chern-Simons level $k$, the central charge of the $\cW^{(2)}_3$ would be reduced by a similar factor.\footnote{As emphasized in \cite{Kraus2}, this can be deduced by replacing the metric in terms of one that gives back the original $AdS_3$ radius but yielding an effective $k/4$, and thus $c$ is reduced by a factor of $1/4$. A more definitive way to compute this is simply to compute the Poisson brackets of the charges generating the asymptotic symmetry transformations. This was done in \cite{Camp:2011}.} 

It is straightforward to make similar statements in the general $N$ case. What one needs to do is to find an explicit representation of the generators of various $SL(2,\bR)$-multiplets, in particular that of the three gravitational spin-one generators in terms of the original ones associated with the principal embedding. Retaining the original metric normalization factor, and denoting the defining vector of the non-principally embedded $SL(2,\bR)$ algebra by $\tilde{L_0}$, we then have the metric describing an $AdS_3$ of radius $R_{AdS_3}$ which changes as
\be
\label{AdS3radius}
\frac{R^2_{AdS_3}}{l^2} =\frac{\textrm{Tr}\left( \tilde{L}_0^2  \right)} {\textrm{Tr} \left( L_0^2\right)}, 
\ee
where $l$ is the radius of the $AdS_3$ vacuum of the principal $SL(2,\bR)$ embedding. 

In \cite{Dynkin,deBoer1}, the general embedding of $SL(2,\bR)$ in $SL(N,\bR)$ was discussed very nicely, and this gives one the basic tools for analyzing this aspect of these higher-spin gravity theories. Following \cite{deBoer1}, let $(n_1,n_2, \ldots)$ be a partition of $N$, with $n_1 \geq n_2 \geq \dots$, then define a different partition $(m_1, m_2, \ldots )$ of $n$, with $m_k$ equal to the number of $i$ for which $n_i \geq k$, and let $s_t= \sum_i^t m_i$. The embedded $SL(2,\bR)$'s generators $(\tilde{L}_0, \tilde{L}_{\pm})$ can then be expressed explicitly as \cite{deBoer1}
\bea
\label{embeddedsl2}
\tilde{L}_+ &=& \sum_{l\geq 1} \sum_{k=1}^{n_l-1} E_{l+s_{k-1},l+s_k}\,,\nn \\
\tilde{L}_0 &=& \sum_{l\geq 1} \sum_{k=1}^{n_l} \left( \frac{n_l +1}{2} - k\right) E_{l+s_{k-1},l+s_{k-1}}\,,\nn \\
\tilde{L}_- &=& \sum_{l\geq 1} \sum_{k=1}^{n_l-1} k(n_l - k) E_{l+s_k,l+s_{k-1}}\,,
\eea
where $E_{ij}$ denotes the matrix with a one in its $(i,j)$ entry and zeroes everywhere else. Applying \eqref{embeddedsl2} to the $SL(4,\bR)$ case where there are 15 generators, we can describe the branching of each non-principal embedding as a direct sum of $(2j+1)$-dimensional irreducible $SL(2,\bR)$ representations $\bigoplus_{j\in \frac{1}{2}\mathbb{N} } n_j \cdot \underline{2j+1}$, where $n_j$ is the degeneracy of the spin-$j$ representation. The branching of the $SL(4,\bR)$ fundamental representation $\underline{15}^{(n_1,n_2,\dots)}$ goes as
\newline
\newline
\noindent
$(1)\,\,\,\underline{15}^{(2,2)} \sim 4 \cdot \underline{3} + 3 \cdot \underline{1}.\,$ Apart from $(\tilde{L}_0, \tilde{L}_{\pm})$, this representation consists of three spin-1 multiplets and three singlets. We compute $\text{Tr}(\tilde{L}_0^2) = 1, R_{AdS_3} = \sqrt{\frac{1}{5}}$.
\newline
\newline
\noindent
$(2)\,\,\,\underline{15}^{(3,1)} \sim \underline{3} + \underline{5} + 2\cdot \underline{3} + \underline{1}\,.$ Apart from $(\tilde{L}_0, \tilde{L}_{\pm})$, this representation consists of two spin-1 multiplets, one spin-2 multiplet and one singlet. We compute $\text{Tr}(\tilde{L}_0^2) = 2, R_{AdS_3} = \sqrt{\frac{2}{5}}$.
\newline
\newline
\noindent
$(3)\,\,\,\underline{15}^{(2,1,1)} \sim \underline{3} + 4 \cdot \underline{2} + 4\cdot \underline{1}\,.$ Apart from $(\tilde{L}_0, \tilde{L}_{\pm})$, this representation consists of four spin-1/2 multiplets, four singlets. We compute $\text{Tr}(\tilde{L}_0^2) = \frac{1}{2}, R_{AdS_3} = \sqrt{\frac{1}{10}}$.
\newline
\newline
We note that in the notations above, the principal embedding is $\underline{15}^{(4)} \sim \underline{3} + \underline{5} + \underline{7}$, and we have set the radius $l=1$. We will use the above results in understanding a subtle aspect of the asymptotic behavior of the black hole solutions in the next Section.

\section{Black holes coupled to Spin-4 current}
\label{sec:black holes}
One fundamental principle to bear in mind in interpreting an ansatz like \eqref{spin4ansatz} in a higher-spin gravity theory is that the higher-spin gauge transformations redefine notions of invariance in gravitational physics. Quantities like event horizons and singularities are gauge-dependent. This was demonstrated and explained carefully in \cite{Kraus1,Kraus2} to be valid for any $SL(N,\bR) \times SL(N,\bR)$ Chern-Simons theory of higher-spin, and the explicit example of the spin-3 case analyzed carefully.\footnote{In \cite{Kraus3}, the corresponding black hole solutions with spin-3 chemical potential in $hs[\lambda] \oplus hs[\lambda]$ were found.} 

Before we review important aspects of this beautiful construction and explicitly carry out the spin-4 generalization of the above ideas, we would like to very briefly discuss the role played by Wilson loops in pure 3D gravity (with negative $\Lambda$), and make some preliminary comments about the significance of a holonomy-based approach towards gravitational thermodynamics in 3D higher-spin theories. 

\subsection{Some comments on the role of Wilson loops}
Let us recall the story of pure 3D gravity with $\Lambda<0$ where all bulk solutions are locally $AdS_3$. The 3D Hilbert action is well-known to be equivalent to a $SL(2,\bR) \times SL(2,\bR)$ Chern-Simons theory. In the framework of $SL(N,\bR) \times SL(N,\bR)$ Chern-Simons, where the $SL(2,\bR)$'s can always be embedded, one can, of course, also recover pure gravity, after setting all higher-spin fields to be zero (see for example, Section 2). 

The $SL(2,\bR)$'s are the isometries of $AdS_3$, and apart from the vacuum, one can generate a rich class of non-trivial spacetimes by orbifolding $AdS_3$ by a pair of suitable generators. One particular example is the BTZ solution which derives from the action of two hyperbolic generators. Realizing $SL(2,\bR)\times SL(2,\bR)$ in terms of their left- and right- action on the embedding hyperboloid $X^2+Y^2-U^2-V^2=-1$, we can write
\bea
\label{isometry}
J_1= -\frac{1}{2} \left(  J_{XU} + J_{YV}   \right),\qquad \tilde{J}_1= -\frac{1}{2} \left(  J_{XU} - J_{YV}   \right) \nn \\
J_2= -\frac{1}{2} \left(  J_{XV} - J_{YU}   \right),\qquad \tilde{J}_2= -\frac{1}{2} \left(  J_{XV} + J_{YU}   \right) \nn \\
J_3= -\frac{1}{2} \left(  J_{XY} - J_{UV}   \right),\qquad \tilde{J}_3= -\frac{1}{2} \left(  J_{UV} + J_{XY}   \right) 
\eea
where $J_{ab}\equiv x_b \partial_a - x_a \partial_b$, and $J_i, \tilde{J}_i$ are the generators of the left and right $SL(2,\bR)$ groups. As explained in \cite{BTZ}, the identification group generated by the Killing vector
\be
\label{killing-phi}
\xi_{\phi}=-r_+ ( J_1 + \tilde{J}_1)
\ee 
yields a quotient space which is the static BTZ black hole of radius $r_+$. In the usual BTZ chart, this identification vector is precisely the vector that generates the rotational symmetry $\partial_{\phi}$ of the black hole. Generally, a Killing vector $\xi$ defines a one-parameter subgroup of isometries of $SO(2,2): P \rightarrow e^{t\xi}P$, where $t$ is an integer multiple of $2\pi$. Since the transformations are isometries, the quotient space obtained by identifying points in a given orbit inherits from $AdS_3$ a well-defined metric. 

We can choose to represent the quotient space construction via matrices. Let us write the defining equation of the $AdS_3$ quadric as the condition on the determinant of a matrix $\bf{X}$ as follows
\be
\label{quartic}
\bf{X} = \begin{pmatrix}
V+X & Y+U\\
Y-U & V-X
\end{pmatrix}, \qquad  \textrm{det} |\bf{X}| = 1
\ee
This condition is preserved by a transformation 
\be
\text{\bf{X}} \rightarrow g_l \text{\bf{X}} g_r^{-1}
\ee
 where $(g_l, g_r) \in SL(2,\bR) \times SL(2,\bR)$. The trace is invariant under conjugation of which classes determine different spacetime solutions. For the BTZ, $(g_l, g_r)$ are hyperbolic generators, and in general, it is natural to ask how these matrices can be mapped via a homomorphism to \eqref{killing-phi} (i.e. in this context, we want to represent $e^{t\xi}$ as an action on $\bf{X}$). Relating this to our conventions for $SL(2,\bR)$, we find 
\be
\label{mapping}
L_{-1} = J_3 -J_2, L_1=J_3+J_2, L_0=J_1 
\ee
and similarly for the anti-holomorphic quantities. As an explicit example, consider the static BTZ. One can first parametrize the $SO(2,2)$ group element as 
\be
\label{btzX}
\textrm{\bf{X}} = \begin{pmatrix}
re^{\phi} & e^t \sqrt{r^2-1}\\
e^{-t}\sqrt{r^2-1} & re^{-\phi}
\end{pmatrix}, \qquad ds^2 = -(r^2-1)dt^2 + \frac{dr^2}{r^2-1} + r^2 d\phi^2
\ee
The hyperbolic quotient by $e^{t\xi_{\phi}}$ is realized from \eqref{killing-phi} and \eqref{mapping} as taking\footnote{We can explicitly realize the $SL(2,\mathbb{R})$ generators as $L_0 = -\sigma_z /2, L_{\pm} = (i \sigma_y \pm \sigma_x)/2$.}
\be
g_l = g_r^{-1} = \begin{pmatrix}
e^{tr_+/2} & 0\\
0 & e^{-tr_+/2}
\end{pmatrix}, 
\ee
from which we see that since $t=2\pi\mathbb{Z}$, making a BTZ is equivalent to identifying $\phi$ as a periodic coordinate. Rescaling $r\rightarrow r/r_+$, $\phi \sim \phi+2\pi$ and we can interpret the spacetime as having a horizon at $r_+$, and with ADM mass $\sim r_+^2$. Going once around the non-contractible cycle along $\phi$, we have $e^{\oint_\phi \xi_{\phi}}$ acting on $\textrm{\bf{X}}$ via $(g_l, g_r)$ as demonstrated explicitly. 

Let us now view things from the Chern-Simons perspective. In the $SL(2,\bR) \times SL(2,\bR)$ Chern-Simons theory, the solutions are flat connections and thus locally can always be expressed as pure gauges, i.e. $A=g^{-1}dg$. Globally, when the spacetime has non-trivial topology, then the gauge function $g$ is not single-valued. When the spacetime has a non-contractible cycle $C$, as we go around the cycle once, $g$ attains a factor of the holonomy $\mathcal{P} \textrm{exp} \left( \oint_C A \right)$. Up to an overall gauge transformation, the flat connections are thus uniquely specified by their holonomies around non-contractible cycles of the manifold. 

Indeed, the holonomies (of $(A,\tilde{A})$) are precisely the $(g_l, g_r)$ described above. \footnote{We note that this relationship between identifications and holonomies is emphasized and discussed nicely in Section 1.3 of \cite{carlip}.} For the BTZ, the $\phi$-cycle is non-contractible, and will have a non-trivial holonomy. From \eqref{ads3}, we have checked that the eigenvalues of $\oint a_{\phi} d\phi = 2\pi a_+$ are precisely those of $-2\pi r_+ L_0$, and similarly $\oint \bar{a}_{\phi} d\phi = -2\pi \bar{a}_{-}$ shares identical eigenvalues with $-2\pi r_+ \tilde{L}_0$. Now, there is another time-like Killing vector $\xi_t \sim - \left( J_1 - \tilde{J}_1\right)$. After a Wick rotation, the thermodynamics of the black hole can be obtained by demanding that the $SL(2,\bR)$ connection be single-valued along Euclidean time direction which is periodic. This cycle is contractible and vanishes at the horizon. The periodic identification of the thermal time can be represented via a pair of $(g_l, g_r)$ acting on $\bf{X}$. From the metric in \eqref{btzX}, we compute the period to be $2\pi$. This means $g_l = e^{2\pi L_0}, g_r^{-1} = e^{-2\pi \tilde{L}_0 }$. Like in the case of the $\phi$ direction, this is equivalent to the action of $e^{\oint_{\tau} \xi_{\tau}}$ if we take $\xi_{\tau}=-r_+\left( L_0 - \tilde{L}_0 \right)$ since $\tau=1/r_+$. The eigenvalues are $\pm \pi$ which we check to be equivalent to that of  $\oint a_{\tau} d\tau = 2\pi \tau a_+$ and $\oint \bar{a}_{\tau} d\tau = 2\pi \tau \bar{a}_{-}$. 

For our purpose, it is natural to ponder about whether a similar interpretation for Wilson loops holds for the gravitational thermodynamics in the context of higher-spin spacetime geometries? We can begin to answer this question by first embedding pure gravity, and thus the ordinary BTZ, in the $SL(N,\bR) \times SL(N,\bR)$ theory. After a straightforward computation, we find that the $N$-dimensional fundamental representation of $SL(2,\bR)$ yields the following set of $N$ eigenvalues for the holonomies for any even $N>2$:
\be
\label{eigenvalues}
\left( \pm  (N-1) \pi, \pm (N-3) \pi, \pm  (N-5) \pi,\, \dots \,\pm  \pi\right)\,
\ee
with a similar result for any odd $N$ but with the last pair of values replaced by $0$. In higher-spin theories, the connection is now also valued in other higher-spin generators, and thus the holonomies no longer correspond neatly to the quotienting action by subgroups of isometries. Nonetheless, we can still classify various solutions according to the holonomies' eigenvalues, as explained nicely in \cite{Castro} and \cite{Maloney}. Apart from higher-spin black holes, as shown recently in \cite{Castro}, when the $\phi$-direction is a contractible spatial cycle, demanding the holonomies to be trivial elements of the $SL(N,\bR)$ gives us a discrete set of solutions which are higher-spin generalizations of conical defects in the pure gravity case. 

\subsection{Solutions in $SL(4,\bR) \times SL(4,\bR)$ Chern-Simons}

In \cite{Kraus1}, the bulk field equations for $SL(3,\bR) \times SL(3,\bR)$ Chern-Simons were solved to yield an ansatz that was interpreted to be a generalized BTZ solution which carries a spin-3 charge. These `black-hole' solutions are defined not by a manifest event horizon (which would be inappropriate since the metric changes under a a general spin-3 gauge transformation), but by demanding that the Wilson holonomy along the time-like direction of \eqref{sol_einstein} has eigenvalues identical to that of the ordinary BTZ black hole. This places constraints on the functions $\cL$ and $\cW$ which are, in principle, functions of the Euclidean time $\tau$ and the chemical potential $\varrho=-\tau \chi$, where $\chi$ is the spin-3 charge (similar relations hold for the anti-holomorphic entities). 

Apart from satisfying gauge invariance, such a prescription was argued to be tenable based on the following resulting conditions: (i)the variables $\cL$ and $\cW$ in the ansatz \eqref{spin4ansatz} satisfy a nice integrability condition: 
\be
\label{integrability}
\frac{\partial \cL}{\partial \rho}=\frac{\partial \cW}{\partial \tau}
\ee
(ii)in the limit when the spin-3 field vanishes, the BTZ is recovered smoothly, and (iii)that there exists a gauge in which the solution exhibits a regular event horizon and spin-3 field, both of which are smooth in the Euclidean $(\tau,\rho)$ plane. Such a black hole is then argued to be a saddle point contribution to a partition function of the form 
\be
\label{partition1}
Z=\tr \left( e^{4\pi^2 i (\tau \cL + \rho \cW - \bar{\tau}\bar{\cL} -\bar{\rho}\bar{\cW} )}  \right)
\ee
By performing a Legendre transform, we can compute the entropy from this partition function. This generalizes the usual notion of the area law. On general grounds of symmetry, it may be natural to expect that if the above procedure works fine for the spin-3 case, it is likely to be valid as well for all higher spins, since $\cW_N$ symmetry arises as the asymptotic symmetry of these higher-spin theories as demonstrated in \cite{Camp:2011}. 

But we think that the consistency of such an approach cannot be guaranteed by asymptotic symmetry arguments alone, and it would be important to analyze some manageable cases in the absence of a general and rigorous proof.

In this Section, we will study the case of spin-4 carefully, and demonstrate that the Wilson-loop-defined black holes can be understood precisely in the same elegant fashion as described above. Let us begin by reviewing the general algorithm of constructing higher-spin black holes as instructed in \cite{Kraus3}: (i)begin by computing the flat connection for the ordinary BTZ in the $SL(N,\bR)\times SL(N,\bR)$ theory (ii)extend this connection by including higher-spin charges together with their conjugate chemical potentials (iii)compute the Wilson loop in the time-like direction of the Fefferman-Graham chart and constrain the holonomy's eigenvalues to be identical to those of the BTZ. 

Following \cite{Kraus3}, the eigenvalue constraint equation can be formulated via a system of trace equations:
\be
\label{trace}
\text{Tr} \left(\omega^{n} \right) = \text{Tr} \left( \omega^n_{BTZ}  \right),\qquad n=2,3,\ldots
\ee
Note that we computed these eigenvalues earlier in \eqref{eigenvalues}. In the case of spin-3, \eqref{trace} terminates at $n=3$, and can be solved via a quartic equation. In the general case, a perturbative approach is easier. For example, as briefly mentioned earlier, in a certain $N\rightarrow\infty$ limit, one can lift the algebra of $SL(N,\bR) \times SL(N,\bR)$ to $hs[\lambda] \oplus hs[\lambda]$, and an illuminating perturbative analysis of \eqref{trace} was performed in \cite{Kraus3} for this case. 

For the spin-4 black holes, from the bulk equations, we can straightforwardly compute the connection to be 
\bea
\label{hspin4connection}
A &=& \left(  e^{\rho} L_1 + \cL e^{-\rho} L_{-1} + \cW e^{-2\rho} W_{-2} + \cU e^{-3\rho} U_{-3} \right) dx^{+} + L_0 d\rho \nn \\
&+& \Bigg( \chi \left( e^{2\rho} W_2 + 2\cL W_0 + 2 \cW e^{-\rho} U_{-1} - \frac{24}{5} \cW e^{-\rho} L_{-1} + \left( \cL^2 - 3\cU \right)e^{-2\rho} W_{-2} + 2\cL \cW e^{-3\rho} U_{-3} \right) \nn \\
&+& f \bigg( e^{3\rho} U_3 + 3\cL e^{\rho} U_1 - 6\cW W_0 + \left(3\cL^2 + 3\cU \right)e^{-\rho} U_{-1} + \frac{54}{5} \cU e^{-\rho} L_{-1} - \frac{24}{5} \cW \cL e^{-2\rho} W_{-2} \nn \\
 &&\qquad \left( \cL^3 + \frac{11}{5} \cL \cU - 4\cW^2 \right)e^{-3\rho} U_{-3}   \bigg) \Bigg) dx^-\nn\\
\eea
where the factors of $e^{n\rho}$ for $n=-3\dots 3$ originate from conjugating a $\rho$-independent $a$ with $b=e^{\rho L_0}$. Please note that we have rescaled the variables $\alpha\cL \rightarrow \cL, \beta\cW \rightarrow \cW, \gamma \cU \rightarrow \cU$ in this Section to avoid cluttering the notations, but will emphasize at appropriate moments later on, when the constants $\alpha, \beta, \gamma$ are needed to be restored. 

The other anti-holomorphic $SL(4,\bR)$ connection reads
\bea
\label{spin4connection}
\bar{A} &=& -\left(  e^{\rho} L_{-1} + \bar{\cL} e^{-\rho} L_{1} + \bar{\cW} e^{-2\rho} W_{2} + \bar{\cU} e^{-3\rho} U_{3} \right) dx^{-} - L_0 d\rho \nn \\
&-& \Bigg( \bar{\chi} \left( e^{2\rho} W_{-2} + 2\bar{\cL} W_0 + 2 \bar{\cW} e^{-\rho} U_{1} - \frac{24}{5} \bar{\cW} e^{-\rho} L_{1} + \left( \bar{\cL}^2 - 3\bar{\cU} \right)e^{-2\rho} W_{2} + 2\bar{\cL} \bar{\cW} e^{-3\rho} U_{3} \right) \nn \\
&-& \bar{f} \bigg( e^{3\rho} U_{-3} + 3\bar{\cL} e^{\rho} U_{-1} - 6\bar{\cW} W_0 + \left(3\bar{\cL}^2 + 3\bar{\cU} \right)e^{-\rho} U_{1} + \frac{54}{5} \bar{\cU} e^{-\rho} L_{1} - \frac{24}{5} \bar{\cW} \bar{\cL} e^{-2\rho} W_{2} \nn \\
 &&\qquad \left( \bar{\cL}^3 + \frac{11}{5} \bar{\cL} \bar{\cU} - 4\bar{\cW}^2 \right)e^{-3\rho} U_{3}   \bigg) \Bigg) dx^+\nn\\
\eea
From these connections, we can derive the metric as explained in the discussion surrounding \eqref{field definitions} with the normalization 
\be
\label{norm metric}
g_{\mu\nu} = \frac{1}{\text{Tr} \left(L_0^2 \right)} \text{Tr} \left( e_{\mu} e_{\nu} \right)
\ee
Note that the above is consistent with the choice made in the particular $SL(3,\bR)$ case presented in \cite{Kraus1}. We have checked that this normalization condition corresponds to $AdS_3$ spacetime with unit radius in the limit of vanishing higher-spin charges. Since the difference between the connections is proportional to the vielbein, we can already read off the asymptotic behavior from \eqref{spin4connection}.

For completeness sake, let us display the metric explicitly\footnote{Please refer to the Appendix for our choice of basis for the $SL(4,\bR)$ generators.}:
\bea
\label{spin4metric}
ds^2 &=& d\rho^2 + \frac{1}{5} \left( 2(-3f\cW+\chi\cL)dx^- +2(-3\bar{f}\bar{\cW}+\bar{\chi}\bar{\cL})dx^+ \right)^2 \nn \\
&&+ \frac{12}{5} \left|   (e^{2\rho} \chi + e^{-2\rho} \bar{\cW} )dx^- + (-\frac{24}{5}\bar{\cW} \bar{\cL} e^{-2\rho}\bar{f} + \bar{\chi}e^{-2\rho} (\bar{\cL}^2 - 3\bar{\cU})dx^+   )    \right|^2 \nn \\
&&- \frac{6}{25}  \left| 3f\cL e^{\rho} dx^- + \left( 2\bar{\chi}\bar{\cW}e^{-\rho} + 3\bar{f}e^{-\rho}(\bar{\cL}^2 + \bar{U}) \right)dx^+   \right|^2 \nn \\
&&- \frac{18}{5} \left| (e^{3\rho} f + e^{-3\rho}\bar{\cU}) dx^- + (2\bar{\chi}\bar{\cL}\bar{\cW} + \bar{f}\bar{\cL}^3 + \frac{11}{5} \bar{f} \bar{\cL} \bar{\cU} - 4\bar{f} \bar{\cW}^2  )e^{-3\rho}dx^+  \right|^2 \nn \\
&&- \left| \bar{\cL} e^{-\rho} dx^- + (e^{\rho} -e^{-\rho}\frac{24}{5}\bar{\chi} \bar{\cW} +  e^{-\rho}\frac{54}{5}\bar{f} \bar{\cU}        )dx^+    \right|^2 \,,
\eea
where the notation $\left| \ldots \right|^2$ refers to multiplying the enclosed expression with its conjugate, i.e. all barred quantities become unbarred and vice-versa, and $dx^{\pm} \leftrightarrow dx^{\mp}$. It is straightforward to check that when all higher-spin charges and their conjugate potentials vanish, the metric \eqref{spin4metric} reduces to that of the BTZ with its Noether charges proportional to $\cL, \bar{\cL}$. From \eqref{spin4metric}, we see that as $\rho \rightarrow \infty$, the terms $\sim e^{3\rho}$ dominate, and the metric asymptotes to
\be
\label{asymptotic1}
ds^2 = d\rho^2 - \left( \frac{18}{5} \bar{f} f e^{6\rho} \right) dx^+\, dx^-
\ee
After re-scaling the boundary cylinder, \eqref{asymptotic1} is just global $AdS_3$ with radius = $\frac{1}{3}$. From a holographical perspective, we can say that the addition of spin-4 potentials in the bulk corresponds to adding to the boundary CFT an irrelevant dimension 4 operator, thereby changing its UV behavior. Indeed, from the general algorithm of constructing this class of higher-spin black holes, it is clear that the metric asymptotes to an $AdS_3$ with radius = $\frac{1}{N-1}$, where $N$ is the largest spin added. 

Incidentally, another related point is that for the spin-3 solution in the principal embedding, since the metric asymptotes to an $AdS_3$
with radius $=\frac{1}{2}$, it was interpreted in \cite{Kraus2} as being the $\cW_3^{(2)}$ vacuum. In this case, we observe that clearly, this is not the case for the spin-4 solution in the principal embedding, i.e. \eqref{spin4metric}. The asymptotic $AdS_3$ vacua of solution \eqref{spin4metric} \emph{cannot} be identified with any of those that belong to non-principal $\cW_4$'s, after the metric is normalized to yield an $AdS_3$ of unit radius in the limit of vanishing higher-spin charges. We conclude that this is the case in general for higher-spin black holes of this type, as can be checked from the expression for the defining vector $\tilde{L}_0$ in \eqref{embeddedsl2}. Thus, the identification/interpretation made in this particular aspect for the spin-3 case is due to more of a coincidence.  

In \cite{Kraus1}, the spin-3 `black-hole' solutions are defined by a manifest event horizon by demanding that the Wilson holonomy along the time-like Killing direction of \eqref{sol_einstein} has eigenvalues identical to that of the ordinary BTZ black hole. Such a procedure was shown to be equivalent to \eqref{integrability} which is a necessary condition for the consistency of the gravitational thermodynamics of the solutions. As a bonus, it was checked that the resulting constraints placed on the variables $\cL$ and $\cW$ lead to the solution approaching the BTZ (and its thermodynamical behavior) in the limit of vanishing higher spin and potential. A caveat is that when the holonomy constraint is imposed, the solution does not exhibit an event horizon, but, in the conventional sense, is instead a transversable wormhole connecting two $AdS_3$ vacua. It was further shown in \cite{Kraus2} that there exists a gauge transformation that takes the metric to one in which $g_{tt}$ has a double zero relative to the radial direction, and thus there could be an event horizon. This is reasonable as higher-spin gauge transformations make the usual notions of Riemannian geometry non gauge-invariant, and one is naturally led to proposing gauge-invariant entities like Wilson holonomies to discuss the physics of these solutions. 

In such an approach, that the thermodynamical consistency has been invoked to be the fundamental definition of a `black hole' is clearly motivated by holography. At this point, it is pertinent to recall that in a certain $N\rightarrow\infty$ limit, one can lift the algebra of $SL(N,\bR) \times SL(N,\bR)$ to $hs[\lambda] \oplus hs[\lambda]$. In this limit, and if we add two additional complex scalar fields in the bulk, we arrive at the bulk picture of the Gaberdiel-Gopakumar conjecture. 

Next, we will mainly be concerned about checking the validity of the integrability condition and the interpretation of the gravitational thermodynamics of the spin-4 solutions. We will see that our results demonstrate clearly that the spin-4 generalization of the framework discussed in \cite{Kraus1, Kraus2} is valid, and thus if this program is to work for all $N$, it has, at least, passed an explicit non-trivial check. This is one of the main results of our paper. 

\subsection{Gravitational thermodynamics from the spectrum of the Wilson loops}

Let us begin this subsection by performing an elementary review of some basic thermodynamical relations like \eqref{partition1}. Mainly, this spells out various sign conventions, scaling factors and, of course, to set up the basic interpretative framework.

Since we have two higher-spin fields, \eqref{partition1} should now be
\be
\label{partition2}
Z(\tau, \varrho, \mu, \bar{\tau}, \bar{\varrho}, \bar{\mu}) = \tr \left( e^{4\pi^2 i (-\tau\cL + \varrho\cW + \mu\cU + \bar{\tau}\bar{\cL} -\bar{\varrho}\bar{\cW} -\bar{\mu}\bar{\cU} )}  \right)
\ee
where $\tau$ is the inverse Euclidean temperature of the BTZ in the limit of vanishing higher-spin, and $\varrho=-\tau\,\chi ,\,\,\, \mu=-\tau\,f$ are the chemical potentials\footnote{The rest of this Section will demonstrate why such an interpretation is viable.}. For a static BTZ limit, we can relate the holomorphic and anti-holomorphic quantities to be
\be
\tau=-\bar{\tau},\,\,\,\varrho= \bar{\varrho} ,\,\,\, \mu= \bar{\mu} ,\,\,\,\bar{\cL}=\cL,\,\,\,\bar{\cW}=-\cW,\,\,\,\bar{\cU}=-\cU
\ee
The spin-4 solutions contribute to the above generalized partition function which includes chemical potentials $\varrho, \mu$ conjugate to the spin-3 and spin-4 currents $\cW, \cU$ respectively. From the CFT perspective, assigning the dependence of $\cL, \cW, \cU$ on $\tau, \varrho, \mu$ (and hence generalizing the BTZ) amounts to the following equations for the expectation values:
\be
\label{partition3}
\langle \cL \rangle = \frac{i}{4\pi^2} \frac{\partial \text{ln} Z}{\partial \tau},\qquad \langle \cW \rangle = -\frac{i}{4\pi^2} \frac{\partial \text{ln} Z}{\partial \varrho},\qquad \langle \cU \rangle = -\frac{i}{4\pi^2} \frac{\partial \text{ln} Z}{\partial \mu}.
\ee
This gives us, as necessary conditions, the integrability constraints:
\be
\label{integrability2}
\frac{\partial \cL}{\partial \varrho} = - \frac{\partial \cW}{\partial \tau},\qquad \frac{\partial \cL}{\partial \mu} = - \frac{\partial \cU}{\partial \tau},\qquad \frac{\partial \cW}{\partial \mu} = \frac{\partial \cU}{\partial \varrho}.
\ee
We note that in the spin-3 case, only the first two relations of \eqref{integrability2} are relevant in ensuring  the validity of the Wilson loop prescription. In the following, we will see that all the three equation of \eqref{integrability2} are satisfied, and thus this constitutes an important and non-trivial evidence for the program to hold for the spin-4 case as well. 

In the absence of a geometric area law for the entropy, one can still define it via a Legendre transform of the free energy. Such a definition implies that the first law of thermodynamics will be automatically satisfied, when $\cL$ is identified as the energy-momentum tensor divided by $2\pi$. The entropy $S$ reads
\be
\label{entropy2}
S = \text{ln} Z - 4i\pi^2 \left( -\tau \cL + \varrho \cW + \mu \cU + \bar{\tau} \bar{\cL} - \bar{\varrho} \bar{\cW} - \bar{\mu} \bar{\cU} \right)
\ee
which can be computed once $\cL, \cW, \cU$ are determined. In the spin-3 case, determining these relations from the Wilson loop prescription is still feasible analytically, but it proves to be difficult in spin $>3$ cases. Inspired by the treatment for the $hs[\lambda]\oplus hs[\lambda]$ case \cite{Kraus3}, we shall perform the analysis perturbatively. 

As a warm-up, let us re-visit the spin-3 case \cite{Kraus1}. Absorbing various scaling constants in $\cL, \cW$, the Chern-Simons connection is $A=b^{-1}ab + L_0 d\rho$, with $a$ being 
\be
\label{ansatzspin3}
a_+ = L_1 - \cL L_{-1} - \cW W_{-2},\,\,\,a_- = \chi ( W_2 - 2\cL W_0 + \cL^2 W_{-2} + 8\cW L_{-1})\,,
\ee
for which, upon taking $n=2,3$, \eqref{trace} reads 
\bea
\label{tracespin3}
\frac{-2}{\tau^2} &=& 8\cL - 96\chi \cW + \frac{128}{3}\chi^2 \cL^2, \nn \\
0&=& -\cW + \frac{8}{3}\chi \cL^2 - 16\chi^2 \cL \cW + 64\chi^3 \left( \cW^2 - \frac{2}{27}\cL^3    \right)
\eea
We note that when the higher-spin fields are set to zero, the BTZ Euclidean temperature $\tau$ can be computed by demanding that the Euclidean section is smooth and is related to $\cL$ by $\cL=-1/4\tau^2$. Now, we can expand the variables as
\be
\label{LW}
\cL = \sum_{n} c_n \chi^n \tau^{-n-2},\qquad \cW = \sum_{n} d_{n} \chi^{n} \tau^{-n-3}
\ee
where $n\in \mathbb{Z}^+_0$. Substituting \eqref{LW} into \eqref{tracespin3}, it can be checked that the following recursion relations are solutions:
\be
\label{spin3coef}
c_n = \frac{16(2n+1)}{3(n-1)}\sum_{i+j=n-2}c_ic_j,\qquad d_{n-1}=\frac{4n}{3(n-1)}\sum_{i+j=n-2}c_i c_j
\ee
From \eqref{spin3coef}, we can see that 
\be
\label{spin3coef2}
nc_n= 4(2n+1)d_{n-1}
\ee
Upon restoring the scaling constants $\cL \rightarrow \alpha \cL = \frac{2\pi}{k} \cL, \cW \rightarrow \beta \cW = \frac{\alpha}{4} \cW$, then \eqref{spin3coef2} immediately implies the integrability condition \eqref{integrability}. We note that in the original work of \cite{Kraus1},  the integrability condition was verified by differentiating \eqref{tracespin3} instead. 

We can seek the analogue of \eqref{spin3coef2} in the case of spin-4, and see whether the integrability conditions \eqref{integrability2} hold. The gauge connection was previously displayed in \eqref{spin4connection}, and by setting $\rho=0$, we can obtain $a$ as the simpler yet also valid variable to compute the holonomy's eigenvalues. The three holonomy equations are
\be
\label{spin4trace}
\text{Tr} (a_+ + a_-)^2 = -\frac{5}{\tau^2},\,\,\,\,\,\, \text{Tr} (a_+ + a_-)^3 = 0,\,\,\,\,\,\,\text{Tr} (a_+ + a_-)^4 = \frac{41}{4\tau^4}\,.
\ee
Substituting \eqref{spin4connection} into \eqref{spin4trace}, and expanding the variables as
\be
\label{LWU}
\cL = \sum_{n,m} c_{nm}  \frac{\chi^n\,f^m}{\tau^{n+2+2m}},\qquad \cW = \sum_{n,m} d_{nm} \frac{\chi^{n} f^{m}}{\tau^{n+3+2m}}, \qquad \cU = \sum_{n,m} b_{nm} \frac{\chi^n f^m}{\tau^{n+4+2m}}\,,
\ee
we can express the holonomy equations \eqref{spin4trace} as polynomial equations in the coefficients. For example, the first one (which is the simplest) reads
\begin{align}
\label{spin4coef}
&0=-20 c_{nm} + 144 d_{n-1,m} -288 b_{n,m-1}  -144b_{n-2,m}+ 64 \sum_{
i+j=n-2}\sum_{k+l=m} c_{ik} c_{jl} + 432\sum_{i+j=m-2}\sum_{k+l=n} d_{k,i} d_{l,j} \nn \\
&\,\,- \frac{2496}{5} \sum_{i+q=n-1} \sum_{ j+r=m-1} c_{ij} d_{qr} -\frac{576}{5}\sum_{i+j+k=m-2}\sum_{p+q+r=n} c_{pi}c_{qj}c_{rk} - \frac{1008}{5} \sum_{i+j=m-2}\sum_{p+q=n}c_{pi}b_{qj}. \nn \\
\end{align}
We verify, up to the 5th order, that the following recursive relations analogous to \eqref{spin3coef2} solve \eqref{spin4coef} and the other two holonomy equations in \eqref{spin4trace}
\be
\label{spin4coef2}
-\frac{5}{18}mc_{nm} = (2n+3m+1)b_{n,m-1},\,\,\, \frac{5}{12} nc_{nm} = (2n+3m+1)d_{n-1,m},\,\,\, 3nb_{nm}=-2(m+1)d_{n-1,m+1}
\ee
It is remarkable to note the appearance of the scaling constants $\beta, \gamma$ defined earlier in \eqref{scalingconstants1} in \eqref{spin4coef2}. Indeed, the integrability conditions \eqref{integrability2} are precisely equal to \eqref{spin4coef2} only upon rescaling 
\be
\label{rescale}
\cL \rightarrow \alpha \cL = \frac{2\pi}{k} \cL, \qquad \cW \rightarrow \beta\cW = -\frac{5\alpha}{12} \cW, \qquad \cU \rightarrow \frac{5\alpha}{18} \cU\,. 
\ee
Now recall that earlier, we mentioned in Section 2 that \eqref{rescale} is necessary if we demand that the bulk equations \eqref{eqnfornu}, \eqref{eqnforL} are equivalent to the OPE between the stress-energy tensor and the higher spin operators. In this aspect, it is nice to see that the consistency condition for holography is precisely the one that surfaces when we equate the holonomy condition to the integrability condition. With this compelling evidence, on top of the spin-3 case in \cite{Kraus1}, it is natural to expect this to hold generally for this class of higher-spin gravitational theories.

This strongly suggests that in fact, writing down a valid gravitational thermodynamics for these geometries can be understood as demanding a consistent holographic dictionary, in particular, that the boundary $\cW_N$ symmetry's Ward identities emerge from the bulk equations. 

Let us close this Section by displaying, to some finite order, all the relevant parameters in terms of $\tau$ and the higher-spin chemical potentials. The partition function and entropy can be calculated using \eqref{spin4coef}, \eqref{spin4coef2}, \eqref{entropy2}, \eqref{partition3} to be
\bea
\label{partitionfn}
S&=&4\pi^2 i \left( \frac{1}{2\alpha \tau} - \frac{4\varrho^2}{5\alpha \tau^5} + \frac{27\mu^2}{50\alpha \tau^7} + \frac{63\mu^3}{125\alpha \tau^{10}} + \frac{21\mu \varrho^2}{5\alpha \tau^8}   \right) + \mathcal{O}(4) \\
\label{p2}
\text{ln} Z &=& 4\pi^2i \left( \frac{1}{4\alpha \tau} - \frac{\varrho^2}{5\alpha\tau^5} + \frac{7\mu \varrho^2}{10\alpha \tau^8} + \frac{\varrho^4}{\alpha\tau^9} - \frac{91\mu^2\varrho^2}{50\alpha \tau^{11}} + \frac{9\mu^2}{100\alpha \tau^7} + \frac{63\mu^3}{1000\alpha \tau^{10}}  \right) + \mathcal{O}(5)  \nn \\ 
\eea
We expect \eqref{partitionfn} to furnish the leading order approximation of the partition function of any candidate boundary CFT with $\cW_4$ symmetry. From \eqref{partitionfn}, we observe that when the higher-spin currents vanish, the BTZ entropy, and thus the familiar area law, is recovered. Finally, the curents $\cL,\, \cW, \cU$ read as
\bea
\label{perturbationL}
\cL&=&\cL_0 - \chi^2 \left( 16\alpha \cL_0^2\right) + f^2 \left( \frac{1008}{25} \alpha^2  \cL_0^3 \right)
- f^3 \left( \frac{4032}{25} \alpha^3 \cL_0^4 \right) - \chi^2 f \left( \frac{1792}{5} \alpha^2 \cL_0^3 \right)+ \mathcal{O}(4) \nn \\ \\
\label{perturbationW}
\cW&=&-\chi \left( \frac{8}{3\beta} \alpha^2 \cL_0^2 \right) - \chi f \left( \frac{112}{3\beta}  \alpha^3 \cL_0^3 \right) + \chi^3 \left( \frac{320}{3\beta} \alpha^3 \cL_0^3 \right) - \chi f^2 \left( \frac{5824}{15\beta} \alpha^4 \cL_0^4 \right) + \mathcal{O}(4) \nn \\ \\
\label{perturbationU}
\cU&=&-f \left(  \frac{16}{5\gamma} \alpha^3 \cL_0^3 \right) + f^2 \left( \frac{336}{25\gamma} \alpha^4 \cL_0^4 \right) + \chi^2 \left( \frac{112}{9\gamma} \alpha^3 \cL_0^3 \right) + \chi^2 f \left( \frac{11648}{45\gamma} \alpha^4 \cL_0^4 \right) \nn \\
&&\qquad - f^3 \left( \frac{38528}{125\gamma} \alpha^5 \cL_0^5 \right) + \mathcal{O}(4) \,,
\eea
where $\cL_0 = \frac{1}{4\alpha \tau^2}$ gives us the relation between the $\tau$ and $\cL$ in the BTZ when all higher-spin fields are zero. We note that the perturbative expansions are governed by $\cL, \cU$ having even $\chi$-parity, and $\cW$ having odd $\chi$-parity.  

\subsection{Smoothness of the higher-spin fields}
With the higher-spin parameters $\cL, \cW, \cU$ computed to be \eqref{perturbationL},\eqref{perturbationW} and \eqref{perturbationU} via the holonomy prescription, it is straightforward to check that the resulting solution does not possess an event horizon.  As mentioned earlier, the higher-spin gauge transformations imply that notions like horizons are not gauge-invariant quantities, and the definition of a higher-spin black hole has been based on the holonomy prescription. As explained in \cite{Kraus1} and nicely demonstrated in \cite{Kraus2}, along the gauge orbit of these solutions, one can find an unique  geometry which exhibits a smooth horizon. 

Although in the general case, even for the simplest spin-3 solution, it is tediously challenging (\cite{Kraus2}) to write down the precise gauge that unveils the event horizon explicitly, an easy linearized analysis can already be useful in helping us understand related aspects.  Below, we will allude to the linearized spin-4 solution, and furnish some preliminary evidence that the holonomy prescription yields in some gauge, an Euclidean geometry carrying higher-spin fields, all of which are free of conical singularity at the Lorentzian event horizon. 

If we keep only terms that are linear in the chemical potentials and the higher-spin currents in \eqref{hspin4connection} and \eqref{spin4connection}, then explicitly, the holomorphic connection reads
\bea
\label{linear}
A_{linear}&=&L_0 d\rho + \left( e^{\rho} L_1 + \cL e^{-\rho} L_{-1} + \cW e^{-2\rho} W_{-2} + \cU e^{-3\rho} U_{-3} \right)dx^+ + \nn \\
&+& f\left( e^{3\rho} U_3 + 3\cL e^{\rho} U_1 + 3 \cL^2 e^{-\rho}U_{-1} + \cL^3 e^{-3\rho}U_{-3}  \right)dx^- \nn \\
&+&\chi \left( e^{2\rho} W_2 + 2\cL W_0 + \cL^2 e^{-2\rho} W_{-2} \right) dx^-\,,
\eea
with a similar-looking expression for $\bar{A}$. Keeping to first order, this gives us the BTZ metric since the corrections to the metric are second order and above, with the horizon located at $\rho_{+}=\left( \text{Log} (\sqrt{2\pi |\cL|/k} \right)$. Also, imposing smoothness of the Euclidean geometry in the $(\tau,\rho)$ plane gives us a relation for $\cW, \cU$ in terms of $\chi$ and $\tau$ which correspond to the first terms in \eqref{perturbationW} and \eqref{perturbationU}, as expected. 

For the higher-spin fields defined in \eqref{field definitions} (setting $\lambda = 4$ since we are working in $SL(4,\bR)$ Chern-Simons theory), we define smoothness at the horizon by demanding relations among the field components identical to those determined when we impose the regularity of the $(\tau,\rho)$ plane of the Euclidean BTZ. Explicitly, we demand
\be
\label{smoothness}
\frac{\partial_{\rho}^2\psi_{\phi \phi tt}}{\psi_{\phi \phi \rho\rho }} \bigg|_{\rho_{+}} =
\frac{\partial_{\rho}^2\psi_{\phi tt}}{\psi_{\phi \rho\rho }} \bigg|_{\rho_{+}}= \frac{16 \pi \cL}{k}=\frac{\partial_{\rho}^2 g_{tt}}{g_{\rho\rho }} \bigg|_{\rho_{+}}
\ee
where $\phi$ is the spectator direction, and we are expanding around the horizon. At first order, from \eqref{field definitions}, $\psi_4$ vanishes for the pure BTZ and the smoothness condition for $\psi_4$ is satisfied trivially. 
However, the spin-3 fields are not smooth even at the first order, so a natural question is whether, analogous to the spin-3 case \cite{Kraus1}, one can perform a gauge transformation to let it develop a double zero at the Lorentzian horizon. 

After some experimentation, we simply find that the ansatz that was used for the spin-3 case in \cite{Kraus1} does the job here as well. The spin-3 fields can be computed to be
\be
\label{linearspin3field}
\psi_{\phi \rho \rho} = \frac{16\pi\cL \chi}{k} + \ldots,\qquad
\psi_{\phi tt} = \frac{40\pi \cW}{k} \left( \rho - \rho_{+}  \right) - \left( \frac{60\pi \cW}{k} + \frac{896\pi^2 \cL^2 \chi}{k^2}  \right) 
\left( \rho - \rho_{+} \right)^2 + \ldots
\ee
We then gauge transform on the background connection (that gives purely the BTZ) via a gauge parameter $F(\rho)$, i.e.
\be
\label{gaugetransform}
\delta A = d\lambda + [ A,\lambda ],\qquad \lambda = F(\rho) \left( W_1 - W_{-1} \right) = -\bar{\lambda}
\ee
to obtain
\bea
\label{changeinpsi}
\delta \psi_{\phi tt}&=& -192\sqrt{\frac{2\pi^3 |\cL^3|}{k^3}}  \left( F(\rho_{+}) \left( \rho - \rho_{+} \right)+ 
F'(\rho_{+})(\rho - \rho_{+})^2 \right)  \\
\label{changeinpsi2}
\delta \psi_{\phi \rho \rho}&=&24\sqrt{\frac{2\pi |\cL| }{k}}F'(\rho_+)
\eea
Comparing \eqref{changeinpsi} and \eqref{linearspin3field}, a straightforward calculation shows that if we set
\be
\label{settingF}
F(\rho_+) = \frac{5\cW}{24} \sqrt{\frac{k}{2\pi |\cL^3|}}\,,
\ee
then the term in $(\rho-\rho_+)$ vanishes, removing the previous singularity. Also, upon demanding 
\eqref{smoothness}, we obtain the first term of \eqref{perturbationW}, and thus showing a nice consistency with the holonomy condition. 

\section{Discussion}

In this paper, we have discussed some interesting holographical aspects of three dimensional higher-spin gravity formulated via $SL(4,\bR)\times SL(4,\bR)$ Chern-Simons theory. In particular, we demonstrated explicitly how $\cW_4$ symmetry and the spin-3 and spin-4 Ward identities arise from the bulk equations of motion coupled to spin-3 and spin-4 currents.  Using the recently found technique discussed in \cite{Kraus1,Kraus3}, we constructed an explicit solution that can be interpreted as a spin-$4$ generalization of the BTZ solution. By identifying the eigenvalues of a Wilson loop along the time-like direction of the static BTZ to that of the spin-4 solution, we showed that this yields a remarkably consistent gravitational thermodynamics for the latter. By tracking the scalings of the higher-spin currents, we argued that the consistency condition for a holographic interpretation of the solution is intimately related to the one that arises in the Wilson holonomy prescription. A linearized perturbative analysis to demonstrate smoothness of the higher-spin fields was also performed. We have also briefly discussed the $AdS_3$ vacua that are associated with the non-principal embeddings of $SL(2,\bR)$ in $SL(4,\bR)$ (giving rise to other $\cW_4$ algebras), and noted that the asymptotics of the black hole (in the standard principal embedding) should be interpreted independently of these secondary vacua. 

The platform of our analysis is the elegant program recently presented in \cite{Kraus1,Kraus2, Maloney,Kraus3}, in which the spin-3 case was discussed in great detail, and generalizations to higher-spin cases also being inferred qualitatively. By trying to understand the spin-4 story concretely, we hope that we have furnished an important and non-trivial supporting example for this recent proposal to understand spacetime geometries in higher-spin gravity formulated via $SL(N,\bR)\times SL(N,\bR)$ Chern-Simons theories. Let us end off by mentioning some worthwhile future directions. 

In \cite{Kraus3}, the spin-3 black holes were lifted to solutions in $hs[\lambda]\oplus hs[\lambda]$ Chern-Simons theory. This was basically achieved by adding an infinite series of higher-spin charges and appropriately inserting normalization factors $N(\lambda)$ such that upon truncation of all spins $s>3$, we have the $SL(3,\bR)$ solution with identical generator normalizations. The Wilson holonomy prescription implies that we have to first compute the Wilson loop's eigenvalues for the BTZ via the infinite collection of traces $\text{Tr} \left( \omega^n_{BTZ} \right)\, \forall n \geq 2$, after choosing a trace convention for the lone-star product that reduces correctly to the $SL(3,\bR)$ conventions. The holonomy constraints are then imposed similarly to the ansatz with spin-3 chemical potential. It was demonstrated remarkably in \cite{Kraus3} that this $hs[\lambda]$ solution yields a high-temperature\footnote{The limit taken in \cite{Kraus3} was $\tau, \varrho \rightarrow 0, \varrho/\tau^2$ fixed.} partition function that agrees with that of the boundary CFT at $\lambda = 0,1$ with spin-3 chemical potential inserted. It was argued that the partition function in this limit is shared by the coset minimal model in Gaberdiel-Gopakumar conjecture since these Chern-Simons solutions describe the topological sector of the bulk, and that the results should support the conjecture for other values of $\lambda$\footnote{For $\lambda=1$, the $\cW_{\infty}[\lambda]$ algebra simplifies to the linear algebras $\cW_{\infty}^{\text{PRS}}$ after a non-linear change of basis, and for $\lambda=0$, after projecting out the spin-1 current, one can obtain $\cW_{1+\infty}$ \cite{Pope:1989ew,Pope:1990kc,Bergshoeff:1990yd}. Each of them can then be realized as free bosons and free fermions respectively.} as well.  A natural direction, thus, would be to turn on other higher-spin potentials and if feasible, this may shed further light on the working principles of the duality. For example, for $\lambda=1$, this can be described by a theory of $D$ free complex bosons with central charge $c=2D$, and the spin-4 current reads $U \sim \left(  \partial \phi \partial^3 \bar{\phi} - 3 \partial^2 \phi \partial^2 \bar{\phi} + \partial^3\phi \partial \bar{\phi}   \right)$ \cite{Bakas:1990ry}, and it would be useful to check if this agrees with the gravity result. 

We may also hope to gain a deeper conceptual understanding behind the holonomy prescription. Ideally, more than verifying its validity in specific cases, it would be nice to have a rigorous and general proof of its equivalence to the integrability conditions that ensure a consistent gravitational thermodynamics for these solutions.

Interestingly\footnote{We thank Ori Ganor for bringing this point to our attention.}, in a different scenario, holonomies also seem to play a crucial role when one discusses the entropy function in $AdS_2/CFT_1$ \cite{Sen} which computes the entropy of extremal black holes with a near horizon geometry of the form $AdS_2 \times K$, where $K$ is a compact space, and with  $U(1)$ gauge fields $A^i$ and charges $q^i$. The well-known formula of Sen, $d_{micro}(\vec{q}) = \langle \text{exp}[-iq_i \oint d\tau A^{i}_{\tau} ] \rangle_{AdS_2}$\footnote{The expectation value refers to the path integral over various fields on Euclidean global $AdS_2$ associated with the attractor geometry for charge $q^i$. See, for example, \cite{Sen}.},computes the microstates in the presence of an inserted Wilson loop lying along the boundary of $AdS_2$. Although these two settings are rather different, their geometrical aspects invite a comparison. Note that at constant $\phi$, the BTZ metric reduces to $AdS_2$, and the Wilson loop which we have used to define the black hole also bounds this $AdS_2$ by definition. More remarkably, we note that in both cases, the role played by Legendre transformation is critical in defining entropy in the absence of a bifurcate horizon. For these reasons, it may be interesting to explore this parallel on a deeper level. 

\section*{Acknowledgments}
I am very grateful to Per Kraus, Eric Perlmutter, Andrea Campoleoni, Stefan Fredenhagen, Stefan Pfenninger and an anonymous referee for their valuable comments on a previous draft, and in particular, Ori Ganor, for many explanations and encouragements . I acknowledge support from Berkeley Center for Theoretical Physics during the course of completion of this work.

\appendix
\section{$SL(4,\bR)$ generators}
\label{Appendix}
Below, we collect the fifteen $SL(4,\bR)$ generators which were used to derive the spin-4 black hole solution in the principal embedding. 

\begin{alignat}{6}
& L_0\,=\,\frac{1}{2} 
\begin{pmatrix}
-3 & 0 & 0 & 0\\
0 & -1 & 0 & 0\\
0 & 0 & 1 & 0\\
0 & 0 & 0 & 3
\end{pmatrix} , \qquad
& & L_1 \, = \, 
\begin{pmatrix}
0 & 1 & 0 & 0\\
0 & 0 & 1 & 0\\
0 & 0 & 0 & 1\\
0 & 0 & 0 & 0
\end{pmatrix} , \qquad
& & & L_{-1} \, = \, 
\begin{pmatrix}
0 & 0 & 0 & 0\\
-3 & 0 & 0 & 0\\
0 & -4 & 0 & 0\\
0 & 0 & -3 & 0 
\end{pmatrix} , \qquad \nn \\
& W_0\,=\, 
\begin{pmatrix}
1 & 0 & 0 & 0\\
0 & -1 & 0 & 0\\
0 & 0 & -1 & 0\\
0 & 0 & 0 & 1
\end{pmatrix} , \qquad
& & W_1 \, = \, 
\begin{pmatrix}
0 & -1 & 0 & 0\\
0 & 0 & 0 & 0\\
0 & 0 & 0 & 1\\
0 & 0 & 0 & 0
\end{pmatrix} , \qquad
& & & W_{-1} \, = \,3 
\begin{pmatrix}
0 & 0 & 0 & 0\\
1 & 0 & 0 & 0\\
0 & 0 & 0 & 0\\
0 & 0 & -1 & 0
\end{pmatrix} , \qquad \nn 
\end{alignat}
\begin{alignat}{6}
& W_2\,=\, 
\begin{pmatrix}
0 & 0 & 1 & 0\\
0 & 0 & 0 & 1\\
0 & 0 & 0 & 0\\
0 & 0 & 0 & 0
\end{pmatrix} , \qquad
&& W_{-2} \, = \,12 
\begin{pmatrix}
0 & 0 & 0 & 0\\
0 & 0 & 0 & 0\\
1 & 0 & 0 & 0\\
0 & 1 & 0 & 0
\end{pmatrix} , \qquad
&&& U_{0} \, =\frac{3}{10} 
\begin{pmatrix}
-1 & 0 & 0 & 0\\
0 & 3 & 0 & 0\\
0 & 0 & -3 & 0\\
0 & 0 & 0 & 1
\end{pmatrix} , \qquad \nn \\ 
& U_1\,=\,\frac{1}{5} 
\begin{pmatrix}
0 & 2 & 0 & 0\\
0 & 0 & -3 & 0\\
0 & 0 & 0 & 2\\
0 & 0 & 0 & 0
\end{pmatrix} , \qquad
&& U_{2} \, = \,\frac{1}{2} 
\begin{pmatrix}
0 & 0 & -1 & 0\\
0 & 0 & 0 & 1\\
0 & 0 & 0 & 0\\
0 & 0 & 0 & 0
\end{pmatrix} , \qquad
&&& U_{3} \, =\,
\begin{pmatrix}
0 & 0 & 0 & 1\\
0 & 0 & 0 & 0\\
0 & 0 & 0 & 0\\
0 & 0 & 0 & 0 
\end{pmatrix}, \qquad \nn 
\end{alignat}
\begin{alignat}{3}
& U_{-1}\,=\,\frac{6}{5} 
\begin{pmatrix}
0 & 0 & 0 & 0\\
-1 & 0 & 0 & 0\\
0 & 2 & 0 & 0\\
0 & 0 & -1 & 0
\end{pmatrix} , \qquad
&& U_{-2} \, = \,6 
\begin{pmatrix}
0 & 0 & 0 & 0\\
0 & 0 & 0 & 0\\
-1 & 0 & 0 & 0\\
0 & 1 & 0 & 0
\end{pmatrix} , \qquad
&& U_{-3} \, =-36
\begin{pmatrix}
0 & 0 & 0 & 0\\
0 & 0 & 0 & 0\\
0 & 0 & 0 & 0\\
1 & 0 & 0 & 0
\end{pmatrix}.\qquad 
\end{alignat}

As mentioned in Section 2, these matrices were constructed by starting with $W_2=L^2_+,\,\,U_3=L^3_+$ and then deriving the rest by the lowering operator $L_-$. The $SL(2,\bR)$ generators $(L_0, L_{\pm})$ can be realized via \eqref{embeddedsl2}.

\section{Spin-4 gravitational field equations}
\label{AppendixB}
Below, we write down the full non-linear action and the field equations for spin-4 gravity. In terms of the vielbein-like and the spin connection-like fields, the action \eqref{action} in the case of $N=4$ reads, after some algebra,
\bea
\label{actioninfull}
S & =&  \frac{1}{8 \p G}\int e^a \ww d\o_a + 2e^{abc} \ww d\o_{abc}  + \frac{1}{6l^2}\e_{abc}e^a \ww e^b \ww e^c + \frac{1}{2} \e_{abc} e^a \ww \o^b \ww \o^c  \nn \\
&& + 2e^{ab} \ww d\o_{ab} + \frac{2}{l^2} \e_{abc} e^a \ww e^{b\alpha} \ww e^{c}{}_{\alpha} + 2 \e_{abc} e^a \ww \o^{b\alpha} \ww \o^{c}{}_{\alpha} + 4 \e_{abc} \o^a \ww \o^{b\alpha} \ww e^{c}{}_{\alpha} \nn\\
&& + 2e^{abc} \ww d\o_{abc} + \frac{2}{l^2} \e_{abc} e^a \ww e^{b\alpha\beta} \ww e^{c}{}_{\alpha\beta}
+ 2 \e_{abc} e^a \ww \o^{b\alpha\beta} \ww \o^{c}{}_{\alpha\beta} 
+ 4 \e_{abc} \o^a \ww \o^{b\alpha\beta} \ww e^{c}{}_{\alpha\beta} \nn \\
&& \e_{abc} \bigg\{ 2\o^{a\alpha\beta} \ww \o^{b}{}_{\alpha\gamma} \ww e^{c\gamma}_{\beta} 
+ \frac{2}{3l^2} e^{a\alpha\beta} \ww e^{b}{}_{\alpha\gamma} \ww e^{c\gamma}_{\beta}  
+2\o^{a\alpha} \ww \o^{b\beta} \ww e^{c}{}_{\alpha\beta} \nn \\
&&\qquad+\frac{2}{l^2}e^{a\alpha} \ww e^{b\beta} \ww e^{c}{}_{\alpha\beta} + 4\o^{a\alpha} \ww e^{b\beta} \ww \o^{c}{}_{\alpha\beta}        \bigg\}
\eea
The equations of motion for the gravitational fields are
\bea
\label{EOMspin2}
&& de^{a} +\e^{abc} \o_b \ww e_c + 4 \e^{abc} \left( e_{bd} \ww \o_c{}^d + \o_{b\alpha\beta} \ww e_c{}^{\alpha\beta}\right) =  0 \, , \nn \\
&&d\o^a +  \e^{abc} \left( \frac{1}{2}  \o_b \ww \o_c + \frac{e_b \ww e_c}{2l^2} +
 2\o_{bd} \ww \o_c{}^d + 2\frac{e_{bd} \ww e_c{}^d}{l^2}  +
  2\o_{b\alpha\beta} \ww \o_c{}^{\alpha\beta} + \frac{2}{l^2} e_{b\alpha\beta} \ww e_c{}^{\alpha\beta} \right) = 0.  \nn \\
\eea
from which we note that the higher-spin fields, like torsion fields, destroy the metric compatibility condition. The field equations for the spin-3 fields are
\bea
\label{EOMspin3}
&&de^{ab} + \e^{cd(a|} \left( \o_c \ww e_d{}^{|b)} +  e_c \ww \o_d{}^{|b)} \right)  + \e^{cd(a|} \left( \o_{c\beta} \ww e_d{}^{|b)\beta} +  e_{c\beta} \ww \o_d{}^{|b)\beta} \right)=0, \nn \\
&&d\o^{ab} +\e^{cd(a|} \left( \o_c \ww \o_d{}^{|b)} + \frac{1}{l^2} e_c \ww e_d{}^{|b)} 
+\o_{c\beta} \ww \o_d{}^{|b)\beta} + \frac{1}{l^2} e_{c\beta} \ww e_d{}^{|b)\beta}  \right)=0 \, . \label{eq2}
\eea
Finally, the field equations for the spin-4 fields are
\bea
\label{spin4}
&&de^{abc} =- \frac{2}{3} \e^{ed(a|} \left( \o_{bc)e} \ww e_d +  e_{bc)e} \ww \o_d \right)  - \frac{1}{3} \e^{ef(a|} \left( \o_{b|ek} \ww e^{|c)fk} +  \o_{b|e} \ww e^{|c)f} \right), \nn \\
&&d\o^{abc}=\frac{2}{3} \e^{ef(a|} \left( \frac{1}{l^2} e^{bc)f} \ww e_e +  \o^{bc)f} \ww \o_e  + \frac{1}{4} \left( \o^{b|}_{ek} \ww \o^{|c)fk} +  \o^b_e \ww \o^{|c)f} + e^{b|}_{ek} \ww e^{|c)fk} +  e^b_e \ww e^{|c)f} \right)\right)\nn \\ 
\eea
From these explicit expressions, we can write down precisely the entire set of gauge transformations acting on the higher-spin fields.\footnote{In our derivations above, we have used various trace relations like
$\textrm{tr} \left( J_a J_b J_c\right)=1/4 \e_{abc},\,\textrm{tr} \left( J_a J_c T_{ef}\right)=\eta_{ac}\eta_{fe}+\eta_{af}\eta_{ec},\,\textrm{tr} \left( J _a T_{ef} T_{ij}\right)=\e_{aei}\eta_{fj}$, etc. For the general spin-$N$, the trace relations for the various $J^a$ and $T^{a_1a_2\ldots a_{s-1}}$ are determined by all possible contraction of indices of the generators in the product via the tensors $\eta$ and $\e$.}

\end{document}